%% file: manuscript.tex
\documentclass[preprint2]{aastex631}

\usepackage[T1]{fontenc}
\usepackage{ae,aecompl}
\usepackage{tablefootnote}
\usepackage{graphicx}	
\usepackage{amsmath}	
\usepackage{amssymb}	
\usepackage{color}
\usepackage{newtxtext,newtxmath}
\usepackage{multirow}

\newcommand{\thestar}{PDS~70}

\include{newcommands}

\begin{document}

\title{The Dynamic, Chimeric Inner Disk of \thestar}

\correspondingauthor{Eric Gaidos}
\email{gaidos@hawaii.edu}

\author[0000-0002-5258-6846]{Eric Gaidos}
\affiliation{Department of Earth Sciences\\
University of Hawai'i at M\"{a}noa\\
1680 East-West Road, Honolulu, HI 96822 USA}
\affiliation{Institute for Astrophysics\\
University of Vienna\\
T\"{u}rkenschanzstrasse 17, 1180 Vienna, Austria}

\author[0000-0003-4507-1710]{Thanawuth Thanathibodee}
\affiliation{Institute for Astrophysical Research\\
Department of Astronomy\\
Boston University\\
725 Commonwealth Ave, Boston, MA 02215, USA}

\author[0000-0002-8732-6980]{Andrew Hoffman}
\affiliation{Institute for Astronomy\\
University of Hawai'i at M\"{a}noa\\
2680 Woodlawn Drive, Honolulu, HI 96822, USA}

\author[0000-0001-7664-648X]{Joel Ong}
\affiliation{Institute for Astronomy\\
University of Hawai'i at M\"{a}noa\\
2680 Woodlawn Drive, Honolulu, HI 96822, USA}
\affiliation{NASA Hubble Fellow}

\author[0000-0001-9668-2920]{Jason Hinkle}
\affiliation{Institute for Astronomy\\
University of Hawai'i at M\"{a}noa\\
2680 Woodlawn Drive, Honolulu, HI 96822, USA}
\affiliation{NASA FINESST Future Investigator}

\author[0000-0003-4631-1149]{Benjamin J. Shappee}
\affiliation{Institute for Astronomy\\
University of Hawai'i at M\"{a}noa\\
2680 Woodlawn Drive, Honolulu, HI 96822, USA}

\author[0000-0003-4335-0900]{Andrea Banzatti}
\affiliation{Department of Physics\\
Texas State University\\
601 University Drive, San Marcos, Texas 78666-4684 USA}




\begin{abstract}

Transition disks, with inner regions depleted in dust and gas, could represent later stages of protoplanetary disk evolution when newly-formed planets are emerging.  The \thestar{} system has attracted particular interest because of the presence of two giant planets at tens of au orbits within the inner disk cavity, at least one of which is itself accreting.  However, the region around \thestar{} most relevant to understanding the planet populations revealed by exoplanet surveys of middle-aged stars is the inner disk, which is the dominant source of the system's excess infrared emission but only marginally resolved by ALMA.  Here we present and analyze time-series optical and infrared photometry and spectroscopy that reveal the inner disk to be dynamic on timescales of days to years, with occultation of sub-micron dust dimming the star at optical wavelengths and 3--5 \micron{} emission varying due to changes in disk structure.  Remarkably, the infrared emission from the innermost region (nearly) disappears for $\sim$1 year.  We model the spectral energy distribution of the system and its time variation with a flattened warm  ($T \lesssim 600$K) disk and a hotter (1200K) dust that could represent an inner rim or wall.  The  high dust-to-gas ratio of the inner disk relative to material accreting from the outer disk, means that the former could be a chimera consisting of depleted disk gas that is subsequently enriched with dust and volatiles produced by collisions and evaporation of planetesimals in the inner zone.  


\end{abstract}



\section{Introduction} \label{sec:intro}

CD-40 8434 (\thestar{}), first identified as a T Tauri star based on its infrared excess \citep{Gregorio-Hetem1992}, has emerged as a key object in studies of late-stage protoplanetary disk evolution and planet formation.  The star possesses a highly structured disk with gaps in both dust and gas \citep{Long2018,Keppler2019} within which orbit two giant planets \citep{Keppler2018,Haffert2019}, both of which exhibit signatures of circumplanetary dust and/or accretion \citep{Haffert2019,Christiaens2019,Isella2019,Hashimoto2020,Zhou2021,Benisty2021}.  

\thestar{} also hosts an inner disk interior to the two planets; its $\sim$10 au extent is only marginally resolved by ALMA at sub-mm wavelengths \citep{Keppler2019}.  Dust in this disk dominates the  emission from this system in the mid-infrared (3-25\micron) and spectroscopy by \spitzer{} and \jwst{} find a dust temperature of 400--600K and strong crystalline silicate emission at 10 \micron{} \citep{Perotti2023}.  The disk is gas-poor but not gas-free \citep{Long2018,Keppler2019,Facchini2021,Skinner2022,Portilla-Revelo2023}; CO (and HCO$^{+}$) was detected at mm wavelengths \citep{Portilla-Revelo2023} but not the 4.6-\micron{} fundamental band \citep{Perotti2023} or its 2.3-\micron{} overtone \citep{Long2018}.  \water{} and \cotwo{} have been detected in emission with \jwst{} \citep{Perotti2023}.  Based on the mm-wave CO and continuum emission, \citet{Portilla-Revelo2023} estimated a gas-to-dust ratio of $\sim$10 , compared to the canonical ISM ratio of 100.  Gas could originate in the outer disk and passing through the gap via a ``bridge" suggested by ALMA imaging \citep{Keppler2019}.  Micron-sized dust grains that are dynamically coupled to the gas can be carried into the inner disk, but condensation of silicates and ices at $\sim$30K in the outer disk and trapping of large particles in a pressure bump at its inner edge is predicted to deplete that gas.

Although Balmer H$\alpha$ and X-ray emission from the star lie within the range of non-accreting weak-lined T Tauri stars \citep{Joyce2020} and far-UV continuum emission from accretion shocks is absent \citep{Skinner2022}, the reversed profile of the H$\alpha$ line and fluorescent H$_2$ emission point to primordial disk gas and low-level ($\lesssim 10^{-10}$ \msun\ yr$^{-1}$) accretion \citep{Thanathibodee2020,Skinner2022}.  Absorption in the 1.083 \micron{} He I triplet and emission in the forbidden [O~I] 6300\AA\ line \citep{Campbell-White2023} reveal a wind that could drive this accretion \citep{Thanathibodee2020}.

Like many other T Tauri stars, emission from \thestar{} is variable \citep{Batalha1998}, and this variability is a probe of conditions in the inner disk.  Sources of variability among T Tauri stars include; at optical wavelengths, accretion, rotation of the spotted stellar surface, and occultation by dust \citep{Herbst1994,Cody2014}; at 3--5 \micron, hot dusty structures near the inner disk rim; and at longer wavelengths by time-dependent self-shadowing of the disk \citep{Muzerolle2009}.  In some systems, anti-correlated ``see-saw" variability of infrared emission around a pivot at $\lambda \approx 5-8$ \micron{} is observed.  The mid-infrared continuum of \thestar{} differs significantly in \spitzer{} and \jwst{} spectra obtained $\approx$15 years apart \citep{Perotti2023}.  Photometry of \thestar{} by the \wise{} mission shows $\approx$0.15 mag variability in 25 \micron{} emission over two one-day intervals that is anti-correlated with that at 3.4 and 4.6 \micron.\footnote{Shadowing could also be responsible for asymmetric HCO$^{+}$ emission from the \emph{outer} disk of \thestar{} \citep{Long2018}, but the asymmetry seems conserved between two observations separated by over a year and it is unlikely these are shadows from the inner disk.}  

Here we present time-series photometry and spectroscopy of \thestar{} that more fully reveal its variability, and the nature of its inner disk.  We propose that the variability and its time-dependence is a manifestation of the mode of accretion at the mobile inner edge of the disk, and is driven by variation in the stellar magnetic field and/or disk feedbacks on $\sim$1 year timescales.  We also propose that the inner disk is a ``chimera", consisting of depleted primordial gas mixed with the solid and gaseous products from the collisions, disintegration, and evaporation of planetesimals.

\section{Observations and Data Reduction}
\label{sec:observations}

\subsection{\tess\ photometry}
\label{sec:tess}
The \tess\ mission \citep{Ricker2014} observed \thestar{} during Sectors 11 (23 April to 20 May 2019), 38 (29 April to 26 May 2021), and 65 (4 May to 2 June 2023, Fig. \ref{fig:all}).  Two-minute cadence PDCSAP lightcurves from Sectors 11 and 65 processed by the \tess\ mission and the 10-minute cadence Science Processing Operations Center (SPOC) lightcurve from Sector 38 were retrieved from the Mikulski Archives for Space Telescopes (Fig. \ref{fig:all}).  For each lightcurve, we performed a Lomb-Scargle periodograms \citep{Scargle1982} and computed the asymmetry $M$ and quasi-periodicity $Q$ parameters defined by \citet{Cody2014}. 

To evaluate the photometry stability of \tess\ photometry \emph{between} sectors we identified 12 stars within 20' and 0.5 magnitudes of the \gaia\ $G$-band brightness of \thestar{} that have both Sector 11 and 38 lightcurves available (Sector 65 lightcurves of these stars were not available at that time of this manuscript's preparation) and compared the median PDCSAP values.  The median absolute deviation is 1.2\%, much smaller than the observed variation for \thestar.  

\subsection{LCOGT Photometry}
\label{sec:lco_phot}

We analyzed photometry obtained by the 0.4-m and 1-m telescope network of the Las Cumbres Global Observatory Telescope (LCOGT) as part of the Key Project ``Catch a Fading Star: Probing the Planet-Forming Zones of Circumstellar Disks with LCO" (KEY-2020-007), as well as previous observations from the public LCOGT archive.   1-m and 0.4-m observations were made between 14 May and 23 June 2020 and  15 Feb and 11 May 2023, respectively (Fig. \ref{fig:all}).  Each 0.4-m telescope was equipped with an SBIG 6303 detector with 0".571 pixels and a 29'$\times$19' field of view (FOV), while the 1-m telescopes are equipped with Sinistro 4K$^2$ detectors with 0".389 pixels and have a 26'$\times$26' FOV.  Exposures were obtained through Sloan $g'$ (4760\AA) and $i'$ (7720\AA) filters with integration times of 10/5 sec (1-m) 280/50 sec (0.4-m).\footnote{Observations were obtained through other filters, but $g'$ and $i'$ provide ample wavelength baseline while avoiding variable H$\alpha$ emission at 6563\AA\ and systematics due to detector ``fringing" in the far red.}  Pointings were designed to maximize the number of comparison stars of comparable brightness.

Images were automatically processed (bias removal, flat-fielding) using the {\tt BANZAI} pipeline \citep{McCully2018}; source identification and aperture photometry was performed with {\tt SExtractor} routines \citep{Bertin1996}.  Instrumental magnitudes were calculated from count rates in an optimal elliptical Kron aperture \citep{Kron1980}.    Lightcurves were produced using the procedures described in \citet{Gaidos2022c} and summarized in Appendix \ref{sec:lcogt_data}.  $g-$ and $i-$band lightcurves were combined into reddening-vs-dimming plots by pairing observations obtained within 60 minutes of each other (the typical cadence of our observations was 6 hr).  We performed weighted linear regression of $g-i$ reddening vs. $i$-band magnitude with three-sigma outlier rejection.  We also rejected points that were brighter or fainter than all other points by $>$0.3 mags.  The former filters flares which are frequent on very young, rapidly rotating stars, the latter filters erroneous $g$-band measurements due to confusion with unrelated fainter sources that appears as variation in $g$-band not accompanied by any $i$-band variation, and produce slope-one excursion in a $g$-$i$ vs. $i$-band diagram.  Uncertainties in slope were calculated by fitting $10^4$ Monte Carlo representations of the data constructed by bootstrap sampling with replacement.

\subsection{ASAS-SN photometry}
\label{sec:asas-sn}

ASAS-SN is a distributed ground-based transient survey (e.g., \citealp{Holoien2017}) with five stations of four telescopes in both hemispheres plus two more telescopes at the Cerro Tololo International Observatory (CTIO) in Chile, one at McDonald Observatory in Texas, USA, and one at the South African Astrophysical Observatory in Sutherland, South Africa. Each ASAS-SN telescope is a 14-cm Nikon telephoto lenses with a back-illuminated, 2048$^{2}$ CCD camera with a 4.47$\times$4.47-degree field of view.   A transition from Cousins $V$-band (5420\AA) to Sloan $g$-band (4750\AA) filters occurred over 2017--2018, with a one-year overlap of observations in both passbands.  Currently, ASAS-SN observes the entire visible sky every clear night to a depth of $g = 18.5$ mag \citep{Shappee2014,Kochanek2017,Hart2023}.  For a given visit ASAS-SN takes three dithered, tandem 90-sec exposures. All images are processed by a fully-automated reduction pipeline (\citealt{Shappee2014}). For the photometry used in this work we use the publicly accessible ASAS-SN Sky Patrol version 1 aperture photometry lightcurves\footnote{\url{https://asas-sn.osu.edu/}} \citep{Kochanek2017}.  This tool performs forced aperture photometry at any position of the sky using a 2-pixel radius ($\approx$16\arcsec) aperture in the {\tt IRAF apphot} package.  We note that \thestar{} is near the optimum brightness for ASAS-SN photometric precision (see Fig. 3 in \citealp{Hart2023}).

\subsection{ATLAS photometry}
\label{sec:atlas}

The Asteroid Terrestrial Impact Last Alert System \citep[ATLAS;][]{Tonry2018} utilizes four 0.5-m f/2 Wright Schmidt telescopes, two in Hawai'i on Haleakal\=a and Mauna Loa, one in South Africa, and one in Chile. During typical survey operations, the telescopes obtain four 30-sec exposures per field each night, allowing the survey to cover the observable sky at daily cadence for equatorial targets and a two-day cadence for polar regions. ATLAS operates with two broad filters, the ``cyan" ($c\approx g + r$) filter (5290\AA) and the ``orange" ($o \approx r + i$) filter \citep[6750\AA,][]{Tonry2018}.  We obtained ATLAS light curves of \thestar{} from the ATLAS forced point-spread function (PSF) photometry service\footnote{\url https://fallingstar-data.com/forcedphot/}, and specifically from the reduced images, prior to the subtraction of a reference image. After excluding epochs with poor data quality and rejecting outliers, we combined the intra-night epochs using a weighted mean to obtain deeper limits and more robust detections. Due to the southern declination of \thestar, the effective cadence is low prior to ATLAS’s expansion to the southern hemisphere in January 2022.

\subsection{WISE photometry}
\label{sec:wise}
We obtained multi-epoch photometry produced by the \wise{} Cryogenic \citep{Wright2010} mission and post-cryogenic NEOWISE \citep{Mainzer2011} and NEOWISE Reactivation \citep{Mainzer2014} missions from the NASA IPAC InfraRed Science Archive (IRSA).  \wise{} cryogenic observations were made in four channels (W1--W4, centered at 3.4, 4.6, 12, and 25 \micron), while NEOWISE data includes only channels W1 and W2.  Magnitudes were converted to flux densities using the zero-magnitude fluxes for flat-spectrum objects published in the Explanatory Supplements to release products for the respective surveys \citep{Cutri2013,Cutri2015}.      

\subsection{LCOGT NRES spectroscopy}  
\label{sec:nres}
Optical (3800-8600\AA), high-resolution ($\lambda/\Delta\lambda \approx 53,000$) spectra were obtained with the Network of Robotic Echelle Spectrographs (NRES; \citealp{Siverd2018}) spectrograph on the 1-m node of LCOGT at CTIO between 13 May and 2 June 2023.  Exposure times were 1800 sec, and the median SNR in the vicinity of the H-$\alpha$ line was 3-9 per spectral pixel.  We performed $n=11$ median filtering of each spectrum and then fit a Gaussian to the smoothed version over the wavelength range 6560--6565.5\AA.  The equivalent width (EW) of the H-$\alpha$ line was computed from the Gaussian fits.

\subsection{HARPS spectroscopy}
\label{sec:harps}
Twenty-two optical spectra (3780--6910\AA) of \thestar, obtained with the High Accuracy Radial velocity Planet Searcher (HARPS) spectrograph on the ESO 3.6-m telescope \citep{Mayor2003} between 29 March 2018 and 15 March 2022, were retrieved from the ESO Archive Science Portal.  The spectral resolution is 120,000 and the median SNR in the vicinity of the H$\alpha$ line is 15-30 pix$^{-1}$, with most near 30.   These spectra were processed by the HARPS pipeline\footnote{https://www.ls.eso.org/lasilla/sciops/3p6/harps/manuals.html}.

\begin{figure*}
    \centering
    \includegraphics[width=\textwidth]{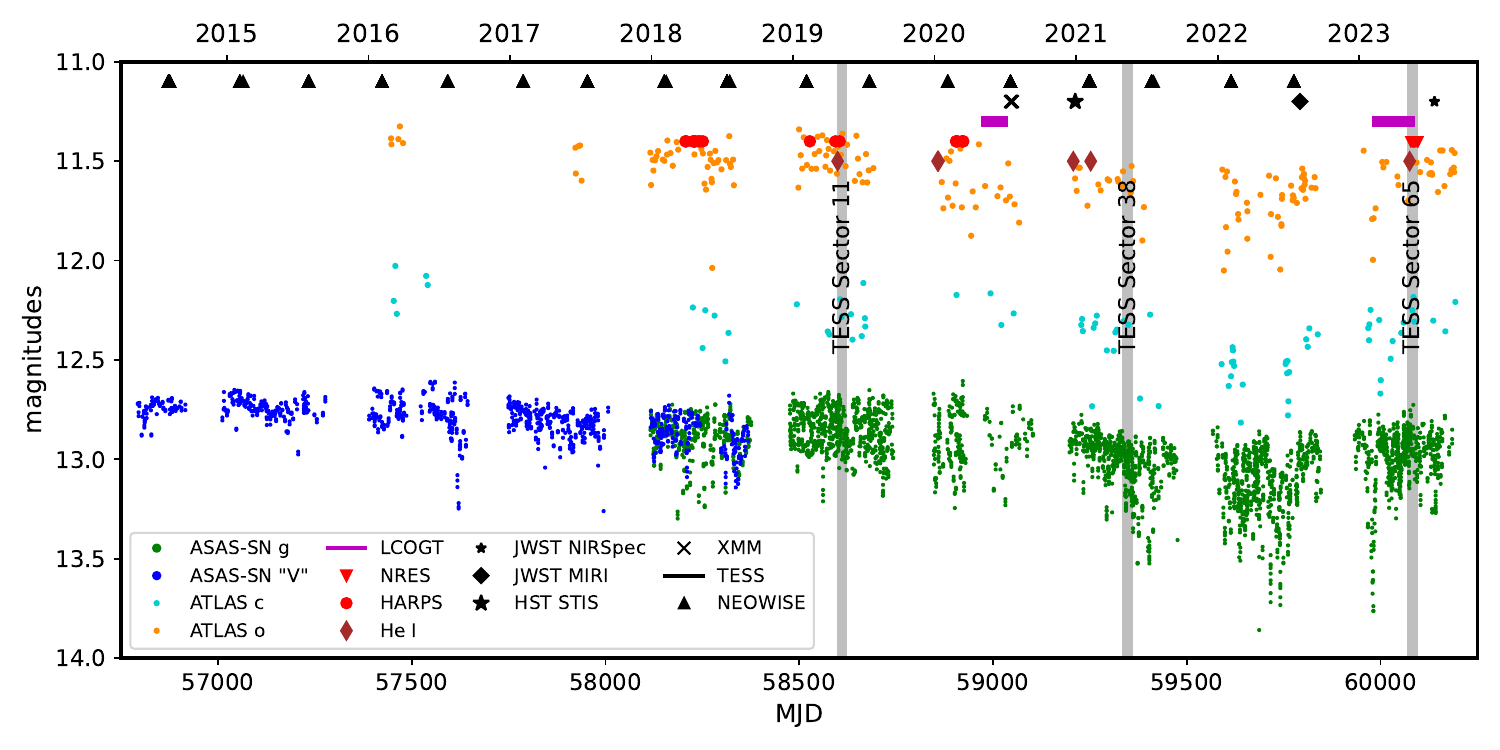}
\caption{Optical lightcurves of \thestar{} by ASAS-SN and ATLAS, plus symbols marking the epochs of other ground- and space-based observations used in our analysis.}
    \label{fig:all}
\end{figure*}

\section{Results}
\label{sec:results}

\subsection{Variability of \thestar{} from circumstellar dust}
\label{sec:variability}

\begin{figure}
    \centering
    \includegraphics[width=\columnwidth]{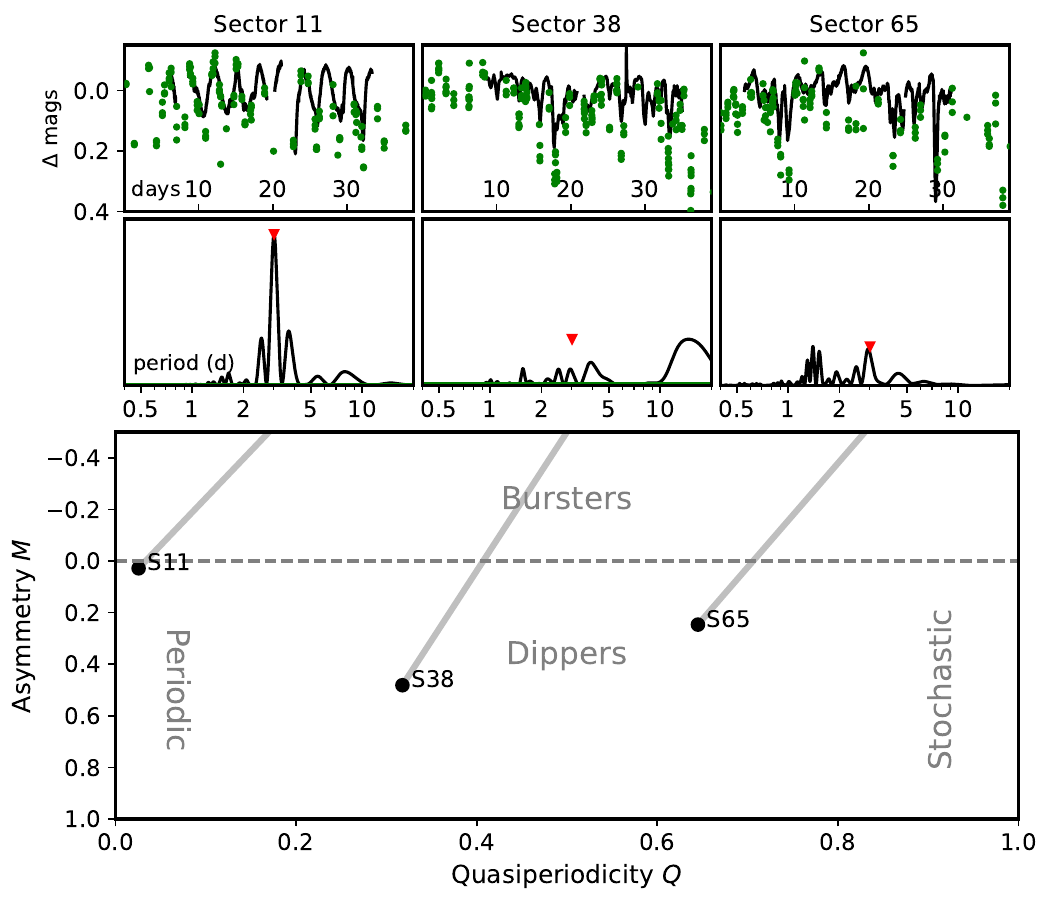}
    \caption{Photometry from \tess{} highlights the evolution of the photometric behavior of \thestar{} over four years.  The top row of sub-panels shows lightcurves in units of relative magnitudes from \tess{} (black lines) during Sectors 11, 38, and 65 as well as ASAS-SN (green points).  The middle row of sub-panels contains Lomb-Scargle periodograms of the \tess\ lightcurves showing the waning signal at 3.03 days taken to be the stellar rotation period (red points).  In the bottom panel the asymmetry $M$ and quasi-periodicity $Q$ parameters \citep[][see text]{Cody2014} derived from the \tess{} lightcurves show a trend from periodic and symmetric (dashed line) behavior to more stochastic and asymmetric (more pronounced dimming) behavior.}
    \label{fig:tess}
\end{figure}

\tess\ lightcurves from three 27-day sectors over 4 years (2019--2023, Fig. \ref{fig:tess}) show photometric behavior that evolved from strongly periodic variability in Sector 11, with a 3.03 day signal previously identified as the stellar rotation period $P_{\rm rot}$ \citep{Thanathibodee2020}, to stochastic, asymmetric variability with weak or no periodicity and dimming events lasting $\sim$1 day in Sectors 38 and 65 (insets of Fig. \ref{fig:tess}).  The bottom panel of Fig. \ref{fig:tess} shows that, after 2019, the lightcurves' quasi-periodicity parameter $Q$ \citep[the fraction of variability not captured by the dominant periodic signal;][]{Cody2014} increases, while their asymmetry parameter $M$ \citep[the skew between the 10th and 90th percentile in the brightness distribution;][]{Cody2014} moves to positive values (more dimming than brightening) characteristic of T Tauri ``dipper" stars that experience transient occultation by circumstellar dust \citep{Cody2014,Ansdell2016a}.

\begin{figure}
    \centering
    \includegraphics[width=\columnwidth]{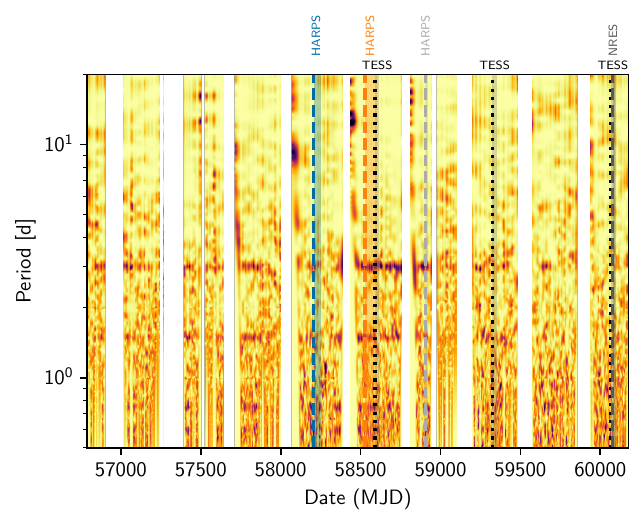}
    \includegraphics[width=\columnwidth]{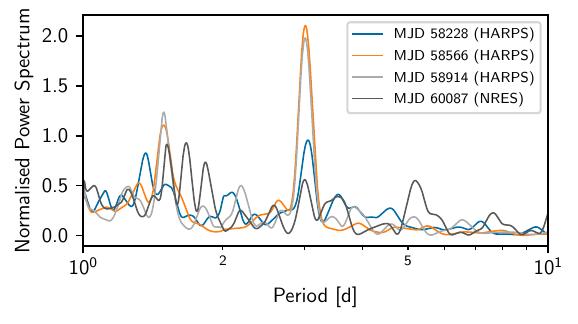}
    \caption{Top: Wavelet-transform power spectrum of ASAS-SN data, shown using a colormap indicating higher power in darker regions. White areas indicate gaps in the ASAS-SN time series. Spectral power at the rotational period of 3 days and its overtones can be seen to emerge starting at MJD 57500, peak at around MJD 58520; and diminish thereafter.   Intervals contemporaneous with observations by \tess{} (Sec. \ref{sec:tess}) and the NRES and HARPS spectrographs (Secs. \ref{sec:tess} and \ref{sec:harps}) are marked by vertical dotted lines.   Bottom: Periodograms constructed from integrating the wavelet power spectrum within the intervals indicated in the top panel, and separately normalised to yield unit integral to permit comparison between temporal baselines of different length.}
    \label{fig:wavelet}
\end{figure}

Both the ASAS-SN and ATLAS photometry show dipper-like variability that increases in amplitude after late 2019, although some of this difference is due to the ASAS-SN change from the $V$- $g$-band filters where scattering by dust (at shorter wavelengths) is greater (Fig.  \ref{fig:all}).  The ASAS-SN data also contain a periodic signal: The upper panel of Fig. \ref{fig:wavelet} shows the Lomb-Scargle wavelet-transform power spectrum of the time-series photometry using using a Gabor-Morlet wavelet, with the parameter governing the trade-off between time and frequency resolution set to $\sigma = 5$. Each band-pass is separately normalised to yield the same median value within the period of overlap.  Spectral power at the rotational period of 3 days, and its overtones, emerges at MJD 57500, peak at around MJD 58520, and diminishes thereafter.

\begin{figure*}
    \centering
    \includegraphics[width=0.49\textwidth]{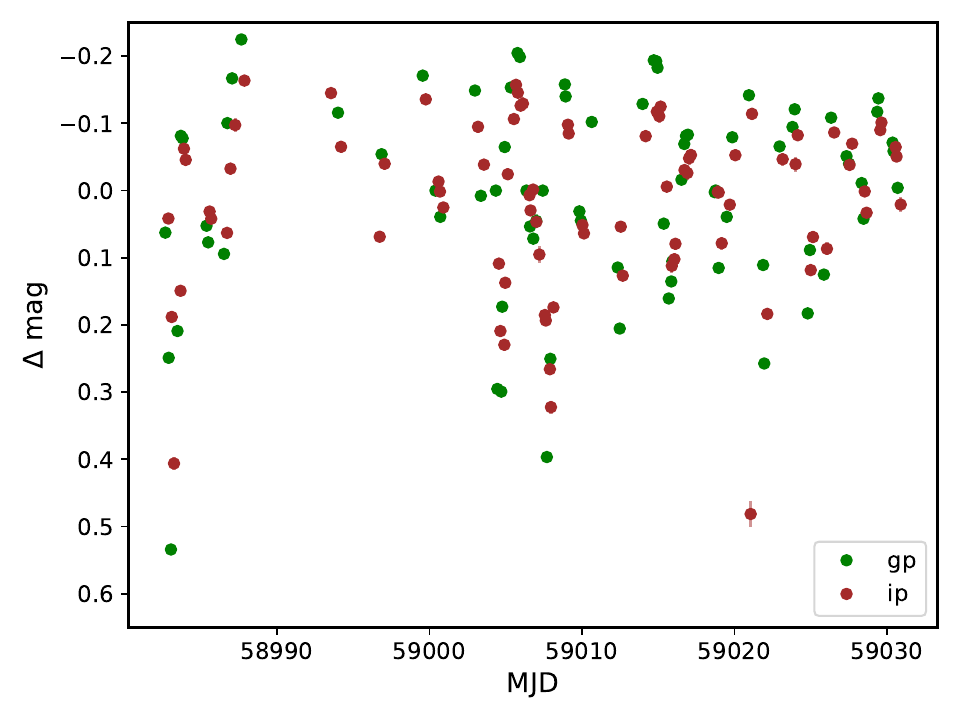}
    \includegraphics[width=0.49\textwidth]{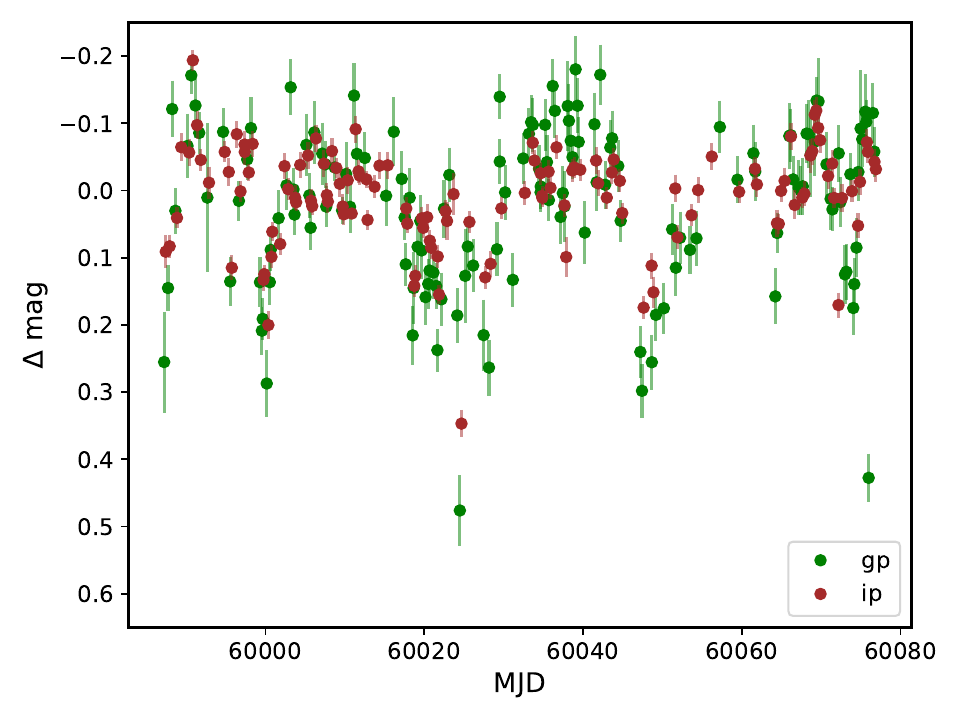}
    \includegraphics[width=0.49\textwidth]{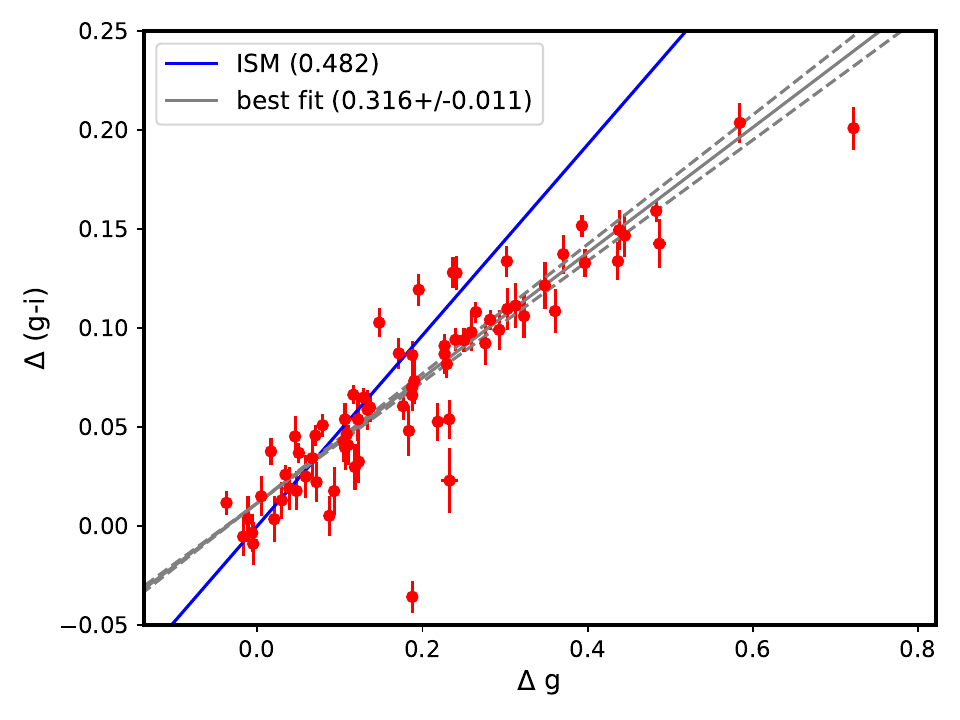}
    \includegraphics[width=0.49\textwidth]{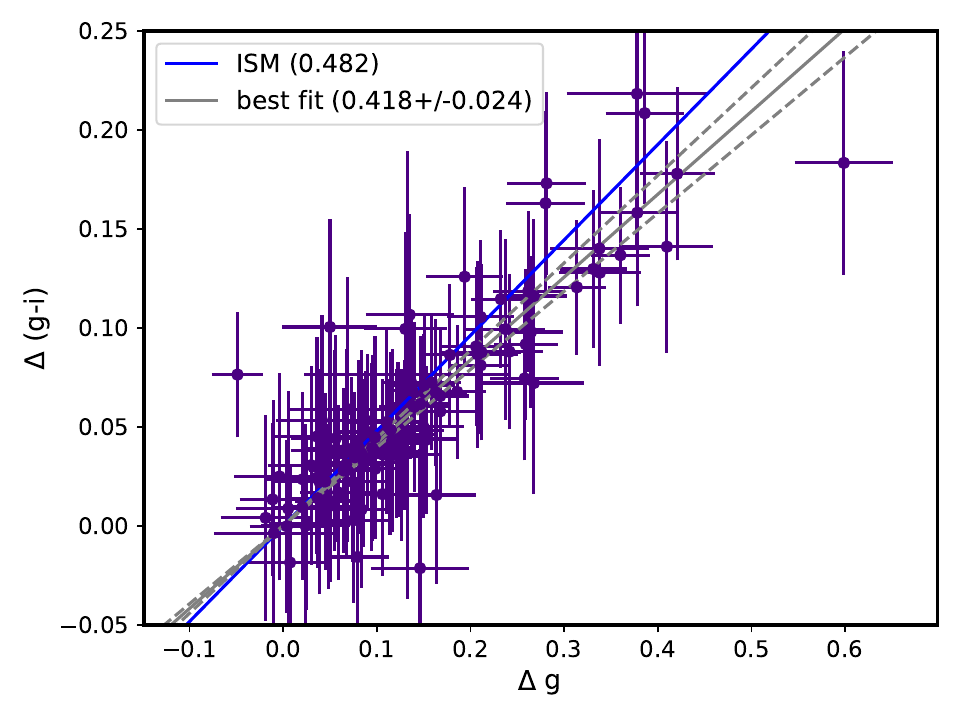}
    \caption{Time-series photometry of \thestar{} obtained with the 1-m (left panels) and 0.4-m (right panels) telescope networks of the LCOGT.  The top panels contain the median-subtracted lightcurves in Sloan $g$- and $i$-bands.  (In the left-hand panel, most errors are smaller than the points).  The bottom panels are reddening vs. dimming plots (in magnitudes), with best-fit lines and the expected interstellar slope calculated based on \citet{Yuan2013} plotted as black and blue lines, respectively.}
    \label{fig:LCOGT}
\end{figure*}

Time-series photometry in Sloan $g$- and $i$-bands obtained in 2020 with the 1-m LCOGT telescope array and nearly three years later with the 0.4-m array also contain dipping behavior (upper panels of Fig. \ref{fig:LCOGT}).  The slopes of the reddening-dimming trends constructed from these lightcurves (lower panels of Fig. \ref{fig:LCOGT}) are consistent with occultation by dust, are much shallower than that seen among YSOs due to variable accretion or flaring ($\approx$0.8; Gaidos et al. in prep), and have greater amplitude than the typical variability produced by rotation and spots .  The best-fit slopes of the two reddening-dimming trends (black lines in in Fig. \ref{fig:LCOGT}) differ significantly and are slightly shallower than the ISM value (blue lines).  

\begin{figure*}
\centering
\includegraphics[width=0.49\textwidth]{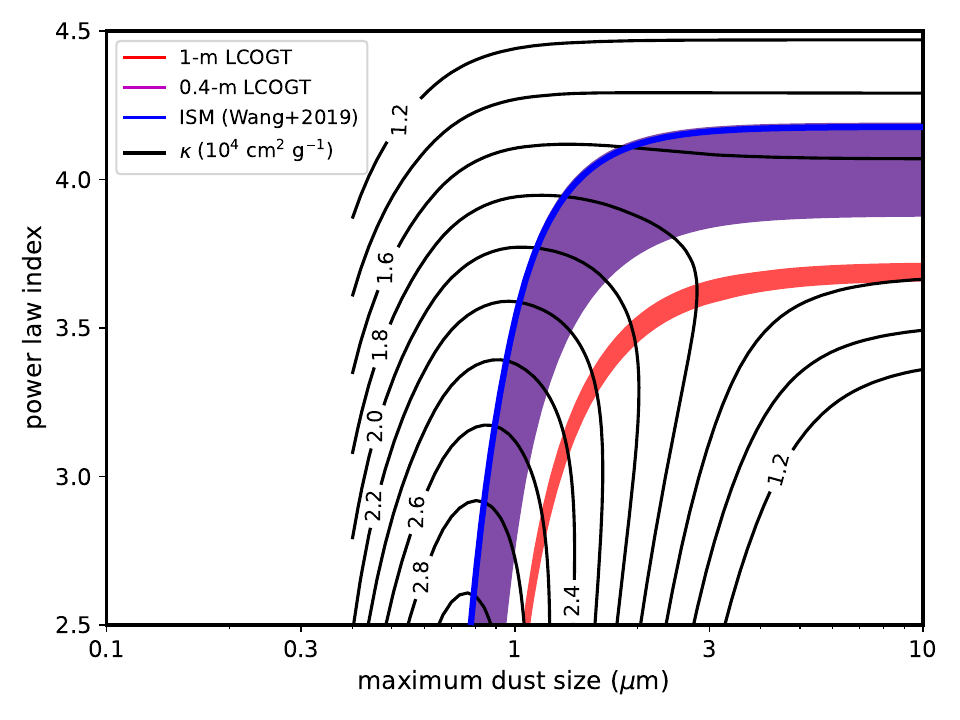}
\includegraphics[width=0.49\textwidth]{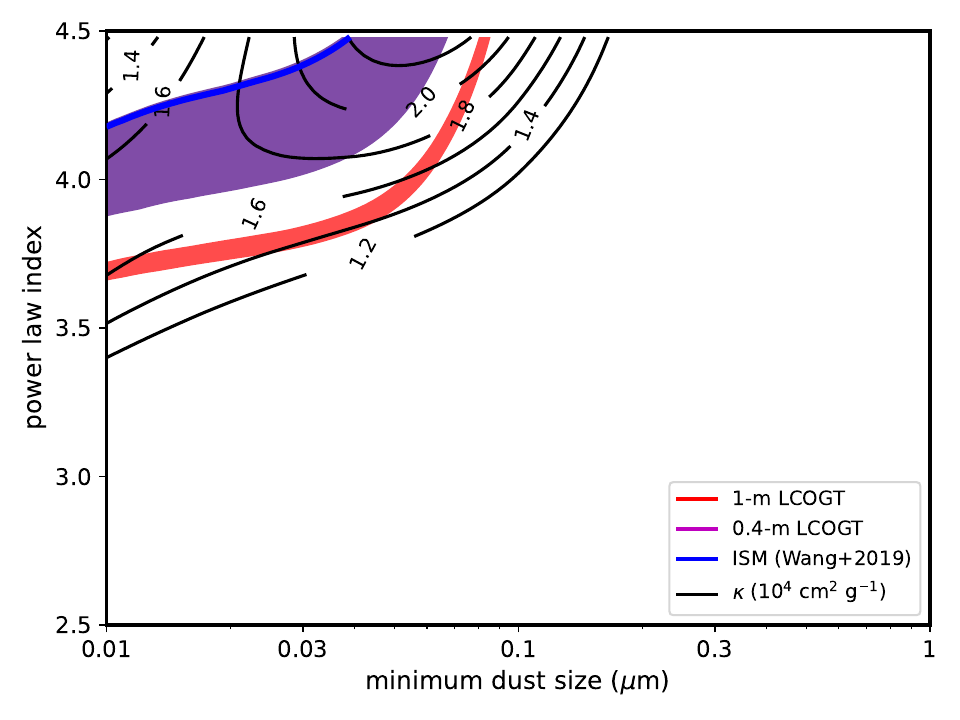}
\caption{Combinations of dust size distribution power-law index (vertical axis) and size cutoff (horizontal axis) for single Mie scattering models that reproduce the $\Delta (g-i)$ vs. $\Delta g$ reddening-extinction slopes obtained from 1-m (red regions) and 0.4-m data (indigo regions) of \thestar{} (bottom panels of Fig. \ref{fig:LCOGT}).  The left panel (a) is for small dust grains with a maximum size (e.g., from grain growth in a protoplanetary disk);  the right panel (b) is for large grains with a minimum size cutoff (e.g. from a collisional cascade in a debris disk).  The blue curve is the ISM and the black contours are of total opacity (scattering + absorption) in units of $10^4$~cm$^{2}$~g$^{-1}$.}
\label{fig:dustsize}
\end{figure*}

We inferred the dust size distribution by comparing these slopes to calculations of absorption and Mie scattering from a power-law particle size distribution with the complex optical constants of \citet{Budaj2015}.  We assumed a pure forsterite composition but the results do not change qualitatively with other silicate-dominated compositions because the extinction is primarily due to scattering.  We used the template spectrum of a K7 solar-metallicity dwarf from the library of \citet{Kesseli2016} and Sloan pass-band response functions from the filter service of the Spanish Virtual Observatory.  We considered two scenarios for the size distribution; one appropriate for the upper layers of a protoplanetary disk which has evolved from an ISM-like distribution where the maximum size of grains is limited by growth vs. settling; and the second appropriate for a debris disk where dust is produced by a collisional cascade and has a power-law distribution with index $p$ close to 3.5 \citep[e.g.,][]{Williams1994,Tanaka1996} and a minimum size is set by Poynting-Robertson drag, radiation blow-out, or gas drag.  Filled contours corresponding to the observed range of reddening-extinction slopes for \thestar{} as well as the ISM are plotted vs. maximum or minimum particle size in Fig. \ref{fig:dustsize}.  Also plotted are contours of total opacity (scattering plus absorption) in units of $10^4$~cm$^{2}$~g$^{-1}$.   Either scenario \emph{requires} the presence of sub-micron grains; the presence of larger grains is allowed, but only if the size distribution is steeper than canonical ($p > 3.5$).

\begin{figure*}
    \centering
    \includegraphics[width=0.49\textwidth]{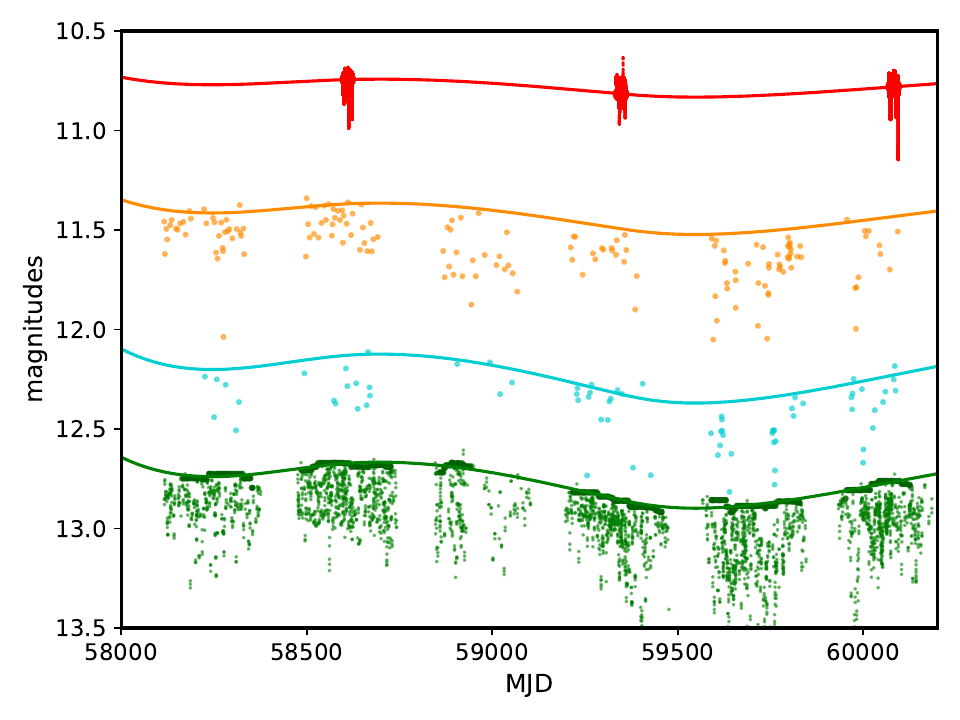}
    \includegraphics[width=0.49\textwidth]{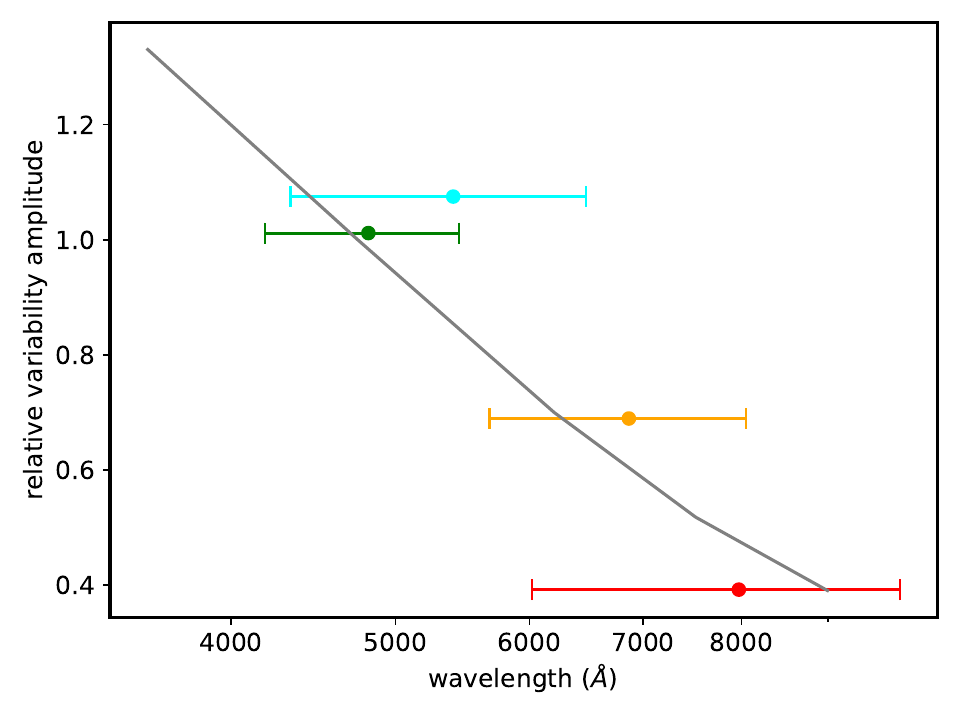}
    \caption{Left (a): From bottom to top: photometry of \thestar{} in ASAS-SN $g$-band (green), ATLAS $c$-band (cyan), ATLAS $o$-band (orange), and \tess\ $T$-band (red).  Solid curves represent a spline curve constructed using the upper boundary of ASAS-SN photometry, and then refit to the other data sets, allowing the amplitude to vary.  Right (b): relative variability in the different passbands vs. wavelength.  The abscissa is the mean wavelength of the passband transmission function, and the error bars span the effective width.  The grey line is the expected trend for ISM-like dust using the extinction coefficients of \citet{Yuan2013}.}
    \label{fig:trend}
\end{figure*}

Rapid (days) variability of \thestar{} is superposed on longer-term variation.  Figure \ref{fig:all} shows there is an envelope of maximum brightness which is most obvious in the ASAS-SN lightcurves, but is also present in ATLAS and \tess\ data.  We fit the envelope of ASAS-SN data by moving a 100-day window over the data, recording the third brightest value with the mean time, and fitting a cubic spline with a smoothing  parameter $s = 0.2$ to those points.  The number of knots in the spline was adjusted by the algorithm so that the $\chi^2$ of the spline fit was $<s$.  We then fit this spline curve to the other, much sparser ATLAS photometry, allowing only the overall amplitude to vary.  In the case of \tess\, we calculated medians of the brightest 90-99 percentile values for each sector and fit these three points.  The inter-sector variation of \thestar{} recorded by \tess{} is 6.8\%, 5.5 times the standard deviation seen among nearby stars (Sec. \ref{sec:tess}).   The ASAS-SN, ATLAS, and \tess{} traces are plotted as the solid curves in Fig. \ref{fig:trend}a and the relative amplitudes of the curves are plotted vs. the passband wavelength in Fig. \ref{fig:trend}b.  The wavelength dependence is consistent with scattering by ISM-like dust (black curve).

\begin{figure}
    \centering
    \includegraphics[width=\columnwidth]{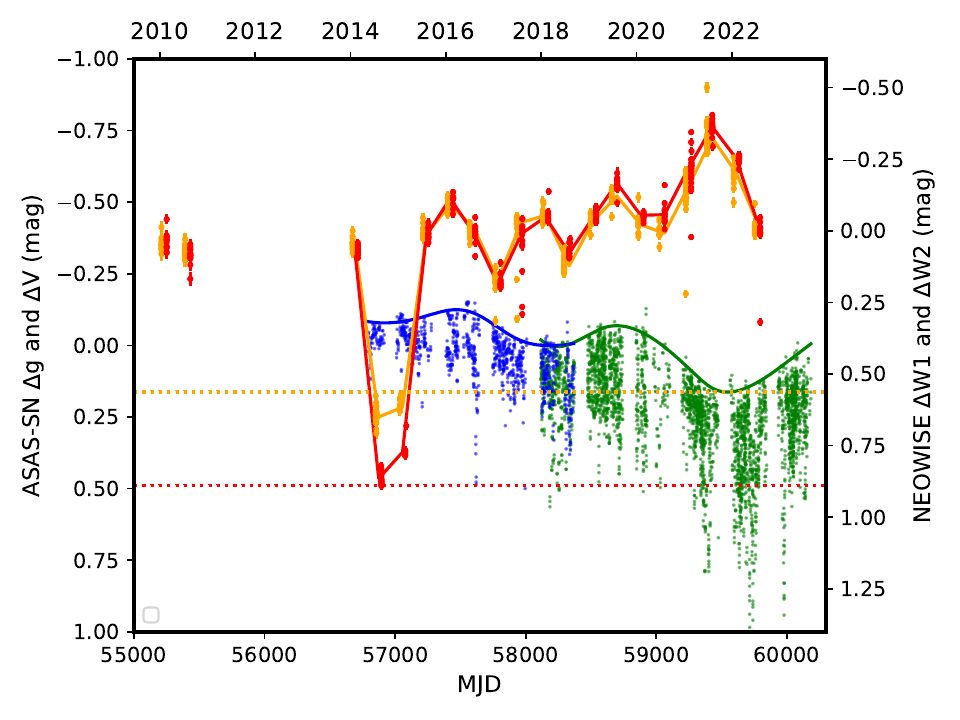}
    \caption{\wise{} 3.4 (W1, orange) and 4.6 \micron{} (W2, red) relative lightcurves of \thestar{} compared to the ASAS-SN $V$-band (blue) and $g$-band (green) lightcurves.  The solid red and yellow lines connect the median values in each biannual visit during the post-cryogenic NEOWISE mission while the earliest, unconnected data are from the Cryogenic mission.  The solid green/blue curve is a spline curve fit to the upper envelope of ASAS-SN photometry. The correspondingly-colored dotted lines indicated the expected disk-less photosphere emission based on the 2MASS $K_s$ (2.2\micron) brightness of the star.}
    \label{fig:wise}
\end{figure}

The cryogenic \wise{} and post-cryogenic NEOWISE 3.4 \micron{} (W1) and 4.6 \micron{} (W2) lightcurves of \thestar{} also contains variation on year-long timescales, with emission appearing to increase as the star dims in the optical (Fig. \ref{fig:wise}).  Most notable is the 0.7--0.9 mag drop in emission for the duration of two consecutive day-long intervals in mid-2014 and early 2015 separated by about 6 months.  The emission in both channels falls to a level close to that predicted for the photosphere flux (horizontal dashed lines in Fi.g \ref{fig:wise}).  This does not appear to be a systematic:  the \wise{} photometric quality flag ({\tt ph\_qual}) is set to "A" and the contamination and confusion flags ({\tt cc\_flags}) are set to zero for all observations, and the $\sim$10\% of observations where frame quality score ({\tt frame\_qual}) is 0 do not correspond to the epochs in question.    Moreover, the lightcurves of 2MASS~J14083163-4121372 and UCAC4~244-066151, the two nearest stars ($\rho$= 4'.6 and 6'.8) of comparable brightness ($\Delta K_s$ = 1.6 and 1.2 mag) show no such dimming during this or any interval of NEOWISE observations.  There is \emph{no} comparable dimming in the ASAS-SN $V$-band lightcurve at these epochs (Fig. \ref{fig:wise}), thus ruling out obscuration of the inner disk (and thus also the central star) by some external occulting object.    This event is exceedingly unlikely to be two or more eclipses (e.g., by an equal-mass binary) because the dimming is consistent over $\sim$2 days, an eclipse longer than this would require a very wide orbit that would be statistically unlikely to be eclipsing.  Also, it is unlikely that NEOWISE would observe two successive eclipses and no others.  Thus the variation must come from disk emission.  Similar drops have been found in the NEOWISE lightcurves of some Herbig Ae/Be stars \citep{Mei2023}.  It is remarkable that after the drop the emission recovers to a level close to that before the event.  

\subsection{Spectroscopic variability due to accretion}
\label{sec:accretion}

\begin{figure}
    \centering
    \includegraphics[width=\columnwidth,angle=0]{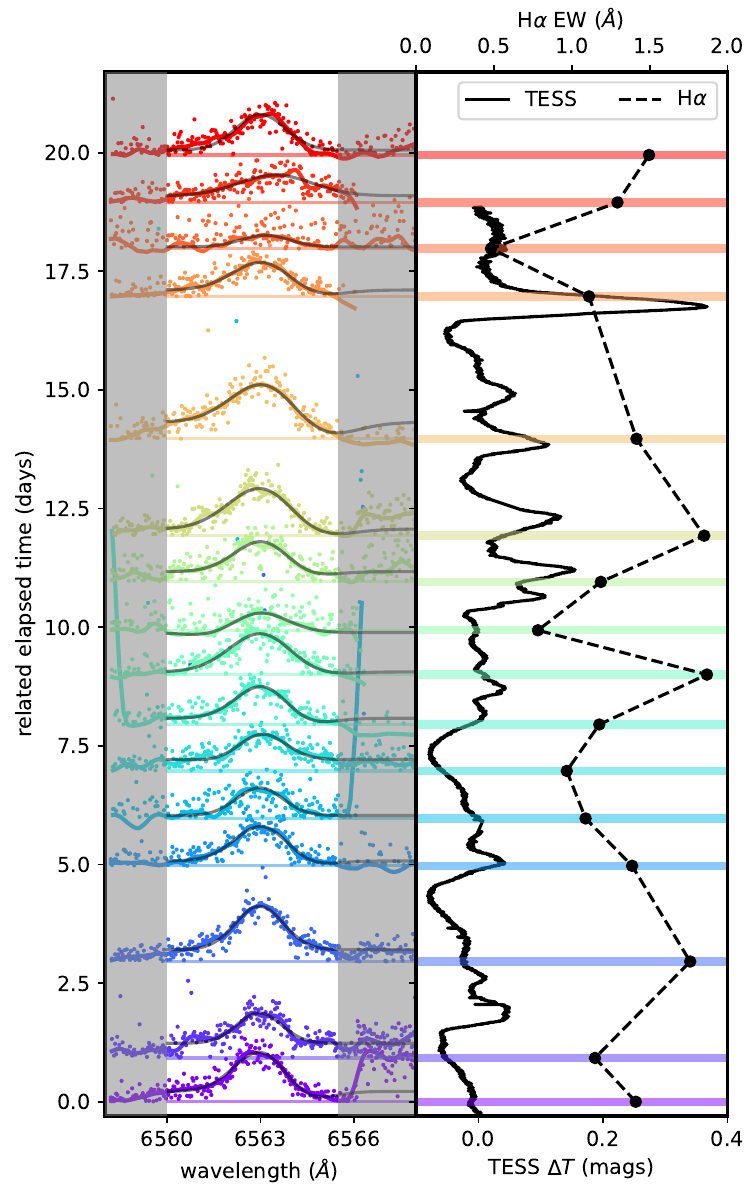}
    \caption{Left (a): Time-series spectra of the Balmer H$\alpha$ line of \thestar{} obtained with the LCOGT NRES spectrograph, with each spectrum placed in observation time relative to the first (obtained on MJD=60077.28).  In addition to the individual data points, an 11-point running median (colored curve) and a best-fit Gaussian (black curve) is plotted.  Points in the grey regions were excluded from the fit.  Right (b): \tess\ Sector 65 lightcurve (solid line) and H$\alpha$ equivalent width from best-fit Gaussians (points and dashed lines).}
    \label{fig:nres}
\end{figure}

NRES spectra contain persistent emission in the Balmer $\alpha$ line of \ion{H}{1}.  The equivalent width (EW) of the line varied between 0.48 and 1.87\AA, characteristic of weak-lined T Tauri stars \citep{Barrado2003}, with no obvious correlation with the optical variability of the star as monitored by \tess{} (Fig. \ref{fig:nres}).  Nor does there appear to be any correlation with stellar rotational phase (adopting a period of 3.03 days) although the phase coverage is limited.   This range of EW overlaps with but is shifted higher than the range found by \citet{Thanathibodee2020} from HARPS spectra obtained in 2018.  

\begin{figure*}
\centering
\includegraphics[width=\textwidth]{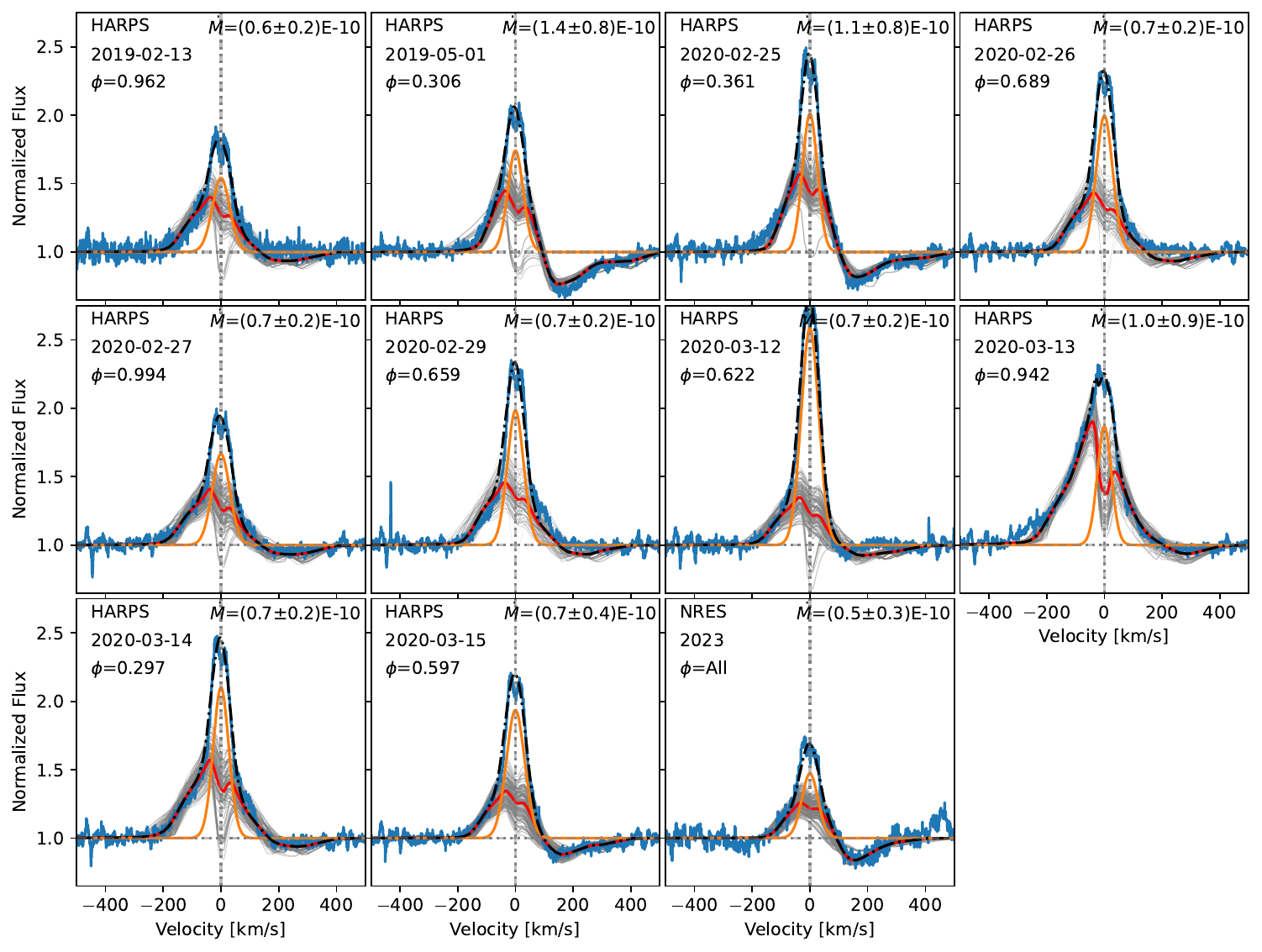}
\caption{Magnetospheric accretion model fit to {\halpha} profiles of \thestar{} obtained with HARPS and NRES. The photosphere-subtracted spectra are plotted in blue. The grey lines are the 100 best fits for a given observation, with the red lines as their averages. The orange lines are the average Gaussian modeling the chromospheric emission, and the black dot-dashed lines are the combined magnetospheric+chromospheric profiles. The observation epochs, rotation phases, and mass accretion rates in M$_{\odot}~$yr$^{-1}$ are reported in each panel.}
\label{fig:harps}
\end{figure*}

We adopted the procedure of \citet{Thanathibodee2023} to determine the magnetospheric accretion properties from the detailed profile of the {\halpha} line.  This procedure is appropriate for weak-lined T Tauri stars and low accretors like \thestar{} where stellar chromosphere makes a significant or dominant (and variable) contribution to the total {\halpha} emission.  Methods based on line luminosity alone \citep[e.g.,][]{Alcala2014} do not disambiguate chromospheric and accretion emission and cannot be reliably applied \thestar, especially since its {\halpha} EW is comparable to that of non-accreting ``template" stars.  We fit each HARPS profile, plus the summed low-signal NRES profiles, with a grid of magnetospheric accretion flow models generated by the code of \citet{Muzerolle2001}. The parameters of the model and their ranges are described in detail by \citet{Thanathibodee2020}. These include accretion rate ($\dot{M} = 0.2-4.5\times10^{-10}\,\msunyr$), maximum flow temperature $T_{\rm max} = 1.0-1.2\times10^4$\,K), magnetospheric truncation radius $R_m = 2-6$\,R$_{*}$), accretion flow width $W_{\rm r} = 0.2-2.0$\,R$_{*}$), and the inclination of the magnetic axis $i$ ($30-75\degr$).  The last is nominally the same as the known stellar inclination ($\approx$50 deg) at each epoch, but was allowed to vary in case the stellar rotation and magnetic axes are not aligned.  The model line profiles were convolved by the instrumental resolutions before fitting. As in \citet{Thanathibodee2023}, we assumed that the cores of the lines are dominated by chromospheric emission, as expected for the low accretion rate of \thestar, and we modeled emission with a Gaussian profile.  We determined the parameters by non-linear least-squares fitting the combined Gaussian and model profile with the {\tt Python Scipy curve\_fit} function, which uses the Trust Region Reflective algorithm \citep{Branch1999}.   The best-fit parameters of a given observation were taken to be the mean form the 100 fits with the lowest $\chi^2$.  Figure~\ref{fig:harps} shows the the {\halpha} lines observed by HARPS at each epoch plus the summed NRES observations (blue lines), along with the best-fit models  The inferred accretion rate and the relative rotational phases (using a 3.03-day rotation period and arbitrary zero point) are reported.  The model does not reproduce some features on a few nights due to the assumed simple geometry (zero-obliquity, axisymmetric, dipolar flow), whereas real accretion flows may be much more complex, as predicted by MHD simulations and suggested by observations \citep[e.g.,][]{Romanova2003,Romanova2021,Zhu2024,Pouilly2021}. The red-shifted absorption is particularly sensitive to the geometry, especially at low accretion rates when flows are expected to have low density and low optical depth \citep{Thanathibodee2023}.

\section{Synthesis and model of the inner disk}
\label{sec:model}

\subsection{Nature of the optical variability}
\label{sec:optical}

Any model for the inner disk of \thestar{} should explain (1) the rapid optical variability due to sub-micron dust that evolves between periodic and aperiodic/stochastic behavior; (2) the long-term variability in both the baseline optical brightness and 3-5 \micron{} emission, also due to dust (although not necessarily the same dust), including the disappearance of the latter for up to one year; and (3) a persistently low but not significantly variable rate of accretion based on H$\alpha$ line profiles.  It should also explain (4) the presence of primordial H and He, as evidenced by UV H$_2$ line emission \citep{Skinner2022}, a reversed H$\alpha$ profile, and a reversed He~I profile \citep{Thanathibodee2019}; and (5) a low gas-to-dust ratio based on sub-mm CO and continuum emission \citep{Long2018}, and the presence of \cotwo{} and \water{} \citep{Perotti2023}. 

\begin{figure}
        \centering
    \includegraphics[width=\columnwidth]{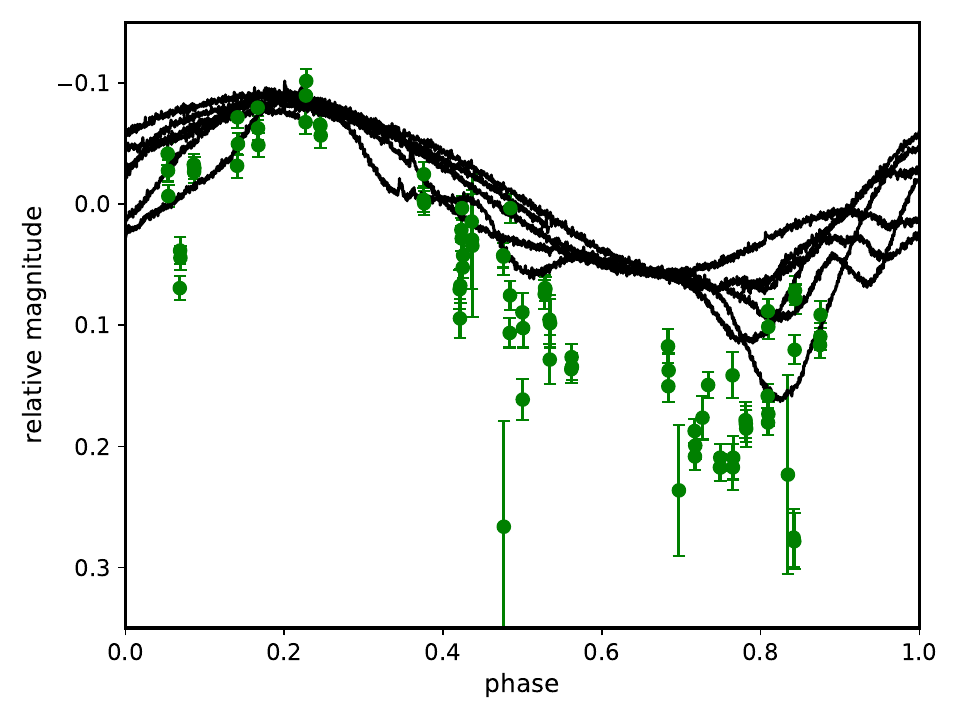}
    \caption{\tess\ Sector 11 lightcurve (black lines) and concurrent ASAS-SN Sloan $g'$ photometry (green points) of \thestar{} phased to the 3.03-day rotation period showing the ``scalloped" morphology and the larger amplitude variation in the shorter-wavelength ASAS-SN photometry.}
    \label{fig:phased}
\end{figure}

\thestar{} is optically variable on timescale of hours to days and its variability has evolved from between purely periodic to aperiodic ``dipper"-like behavior over the timescale of years (Figs. \ref{fig:tess} and \ref{fig:wavelet}).  While the periodic behavior during \tess{} Sector 11 suggests rotational variability by spots, its large and wavelength-dependent amplitude ($\sim$0.2 mags peak-to-peak in the \tess\ passband and larger in ASAS-SN $g'$ data) and ``scalloped" lightcurve morphology (Fig. \ref{fig:phased}) is similar to that seen around other young low-mass stars.  This morphology been explained by scattering by sub-micron dust trapped by the stellar magnetic field at the co-rotation radius, where the orbital period equals the stellar rotation period \citep{Stauffer2017,Stauffer2018,Zhan2019,Guenther2022}.  In contrast, the later aperiodic variability of \thestar{} suggests occultation by dust located exterior to the co-rotation radius \citep{Donati2019}.  The periodic signal (and its upper harmonic) was already present in ASASA-SN data by May 2014, peaked in intensity for about two years beginning mid-2018, then decreased in intensity and disappearing by March 2021, suggesting changes in the innermost location of dust relative to the co-rotation radius.  Our multi-band LCOGT photometry unambiguously indicates that scattering by sub-micron dust is responsible for this dimming (Fig. \ref{fig:LCOGT}), at least during the most recent aperiodic phase.  Superposed on this pattern of rapid variability is an envelope of smoother variation in brightness on the timescale of $\sim$1 year (Fig. \ref{fig:trend}a).  The trend of decreasing amplitude with increase wavelength (Fig. \ref{fig:trend}b) suggests that this also is due to sub-micron circumstellar dust.  Such long-term variability is not highly unusual around T Tauri stars \citep[e.g.,][]{Grankin2007,Rigon2017}.

\citet{Thanathibodee2019} combined the 3-day period from the \tess\ Sector 11 lightcurve with a rotational line broadening of $v \sin i = 16$ km~sec$^{-1}$ from high-resolution spectra to estimate a stellar inclination of $51 \pm  8$ deg, consistent with the inclination of the ALMA-resolved outer disk \citep[$51.9\pm0.2$ deg;][]{Keppler2019}.  If the inner disk is aligned with stellar rotation, then disk dust will intercept the line of sight to the star only if it is diverted well above the midplane, e.g., by the stellar magnetic field.  This can occur where the disk gas is (partially) ionized, magnetic field pressure dominates over viscous stresses or ram pressure produced by accretion, and if dust is dynamically coupled to the gas.  This diversion truncates the disk, and if this truncation occurs inside the co-rotation radius then disk material loses angular momentum to the star and magnetospheric accretion will occur, i.e. through ``funnels" onto the star \citep{Bouvier2007b}.  If truncation occurs outside the co-rotation then some of the disk can be accelerated away from the star and above the mid-plane in a magnetohydrodynamic wind, with a remainder accreted onto the star in ``propeller accretion".  We discuss the potential mechanisms and timescale of this evolution in Secs. \ref{sec:synthesis} and \ref{sec:timescales}. 

\subsection{Nature of the infrared variability}
\label{sec:infrared}

\begin{figure}
    \centering
    \includegraphics[width=\columnwidth]{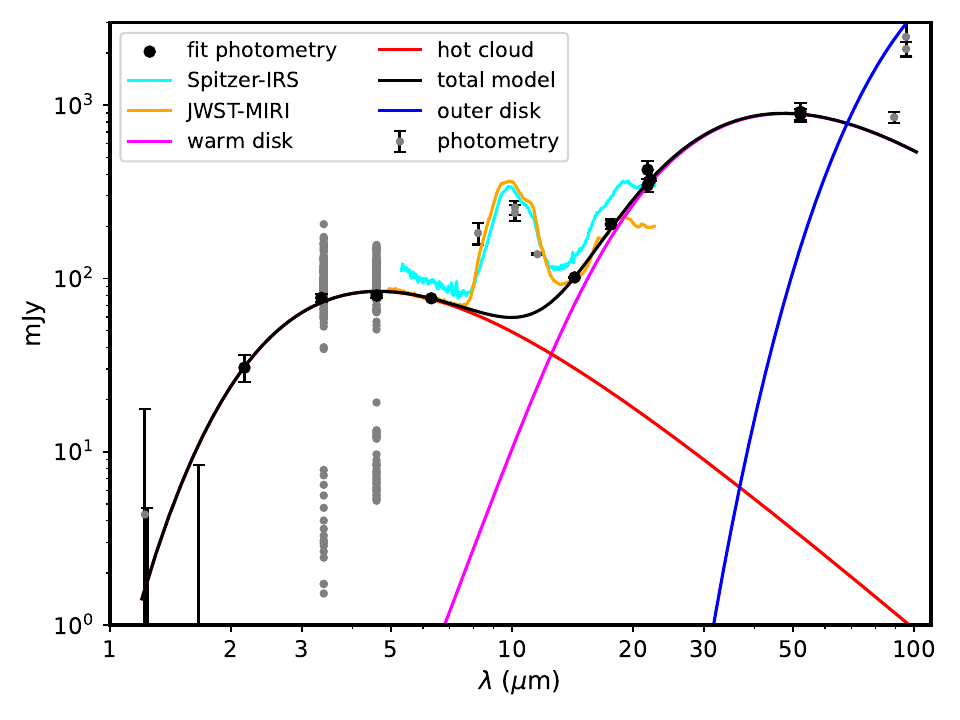}
    \caption{Photosphere-subtracted infrared photometry and spectroscopy of \thestar{} compared to a best-fit SED model (black curve) consisting of an inner disk rim (red curve), and a power-law disk (magenta curve).  Black points are those used in the fit.  The blue curve is the expected emission from the outer disk starting at $\approx$50 au.  This is the dominant contribution to the 100 \micron{} point, which is excluded from the fitting.  The grey points at 3.4 and 4.6 \micron{} are the (excluded) NEOWISE photometry.}
    \label{fig:sed_fit}
\end{figure}

To describe the infrared emission and its variability in terms of this scenario we constructed a spectral energy distribution (SED) of the inner disk from photometry cataloged by the Virtual Observatory SED Analyzer, fit a solar-metallicity BT-SETTL stellar photosphere model \citep{Allard2012} with CFIST solar abundances \citep{Caffau2011} at wavelengths $\lambda \le 1.6$ \micron{} (where the disk contribution is negligible), and subtracted the best-fit model (\teff=4000K, $\log g = 4.5$, $\chi^2 = 0.22$) from the photometry at longer wavelengths (Fig. \ref{fig:sed_fit}).  We then fit a disk model consisting of a ``warm", optically-thick disk with a power-law temperature distribution with semi-major axis, plus an isothermal structure representing a ``hot" interior disk wall or ring.  The second component is motivated in part by the shape of the SED, and by the need for occulting dust close to the star to produce the dipper-like variability; at the 3.03-day co-rotation radius the blackbody equilibrium temperature is $\approx$1500K.  

We fixed the inner disk inclination to that of the outer disk \citep[51.7 deg;][]{Keppler2019}, motivated by the agreement with the stellar inclination \citep{Thanathibodee2020}, leaving five free parameters: hot component area $A_{\rm hot}$ and temperature $T_{\rm hot}$, outer disk edge $r_{\rm outer}$, temperature $T_{\rm warm}$ at the inner disk edge, and power-law index $\beta$.  For this fit we omitted the NEOWISE data (see below), photometry between 7.7 and 13.1 \micron{} which is affected by silicate emission, as well as IRAS photometry at 100 \micron, which is dominated by emission from the \emph{outer} disk that starts at $\approx$50 au where the blackbody temperature would be 30K (blue curve in Fig. \ref{fig:sed_fit}).  We also omitted DENIS $K$-band photometry which is somewhat higher than 2MASS (see Sec. \ref{sec:future}).  We included mean values in the ranges 5--7.7\micron{}  and 13.1--15.5\micron{} from the \jwst{} MIRI spectrum in lieu of the actual spectrum.  A satisfactory fit ($\chi^2=7.5$, 5 d.o.f.) was found for $T_{\rm hot}=1100$K, $A_{\rm hot} = 2.8 \times 10^{-3}$au$^{2}$, $T_{\rm warm}=600$K, $r_{\rm outer} = 3.3$ au, and $\beta=0.22$, and is plotted as the black curve in Fig. \ref{fig:sed_fit}.

\begin{figure}
    \centering
    \includegraphics[width=\columnwidth]{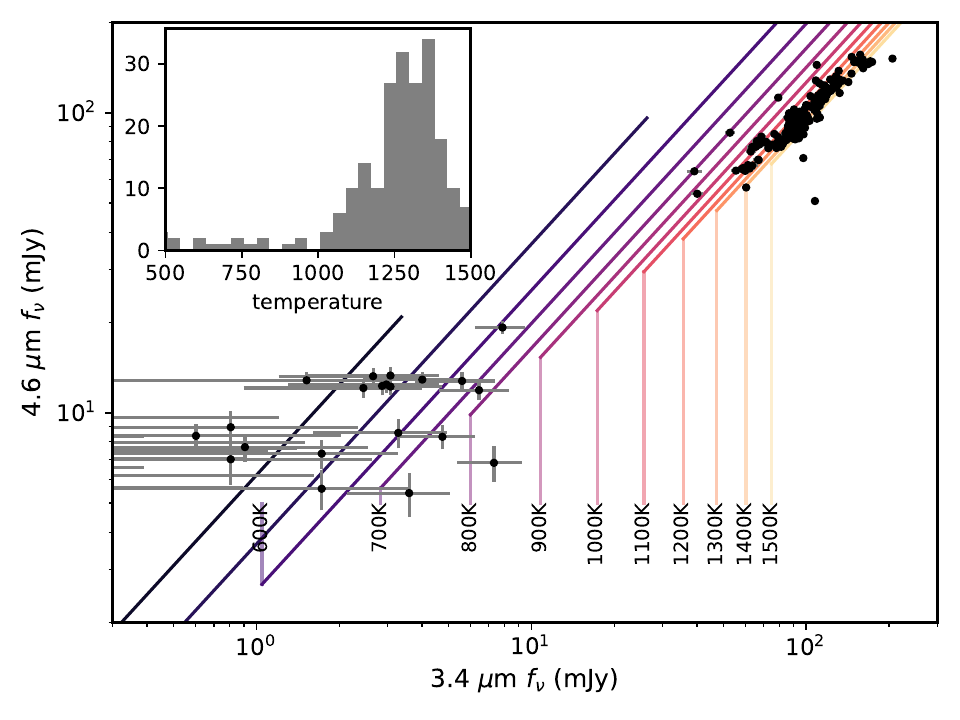}
    \caption{Photosphere-subtracted emission from the inner disk of \thestar{} in the \wise{} W1 (3.4\micron) vs W2 (4.6\micron) passbands (points with error bars), compared with that from black-bodies for different temperatures and emitting areas (lines color-coded by temperature).  The insert shows the distribution of best-fit blackbody temperatures from the observed flux densities.}
    \label{fig:area-temp}
\end{figure}

Variability in the NEOWISE 3.4 and 4.6 \micron{} photometry on day- to year-timescales (Fig. \ref{fig:wise}) implies that the hot dust component must evolve.  We modeled this variability by subtracting the warm disk component, assuming it is not variable or if it changes at all, does not contribute significantly at these wavelengths (Fig. \ref{fig:sed_fit}), and that emission from the hot component can be described as a single but variable temperature.  Figure \ref{fig:area-temp} shows the variation in the W1 and W2 fluxes after the photosphere and small and constant warm disk contribution is removed, as well as curves of blackbody emission for a fixed temperature.  The inset shows the distribution of best-fit temperatures, indicating the relatively narrow dispersion between 1000 and 1500K.  The residual emission (if any) during the low emission state in 2014-2015 is consistent with emission at only $\sim$500K (points in lower-left of Fig. \ref{fig:area-temp}), and thus is likely a result of variation in or imperfect subtraction of the warm disk component.

Some T Tauri stars exhibit variability at longer $\lambda >10$ \micron{} that is anti-correlated with emission at shorter wavelengths; this ``see-saw" variability is thought to arise from shadowing of the disk by a dynamic inner ``wall", which intercepts and re-processes stellar radiation that is otherwise incident on the disk further out \citep{Muzerolle2009,Espaillat2011,Flaherty2012,Fernandes2018}. Indeed, \citet{Perotti2023} found a $\sim$30\% difference in the continuum emission of \thestar{} between \spitzer{} and \jwst{} spectra.  In \wise{} photometry from the cryogenic mission, emission in the W4 (25\micron) channel is significantly anti-correlated with that in W1 and W2 ($p = 0.001$, Spearman rank test) but the variation is small and observations span just two one-day intervals.

\begin{figure}
\includegraphics[width=\columnwidth]{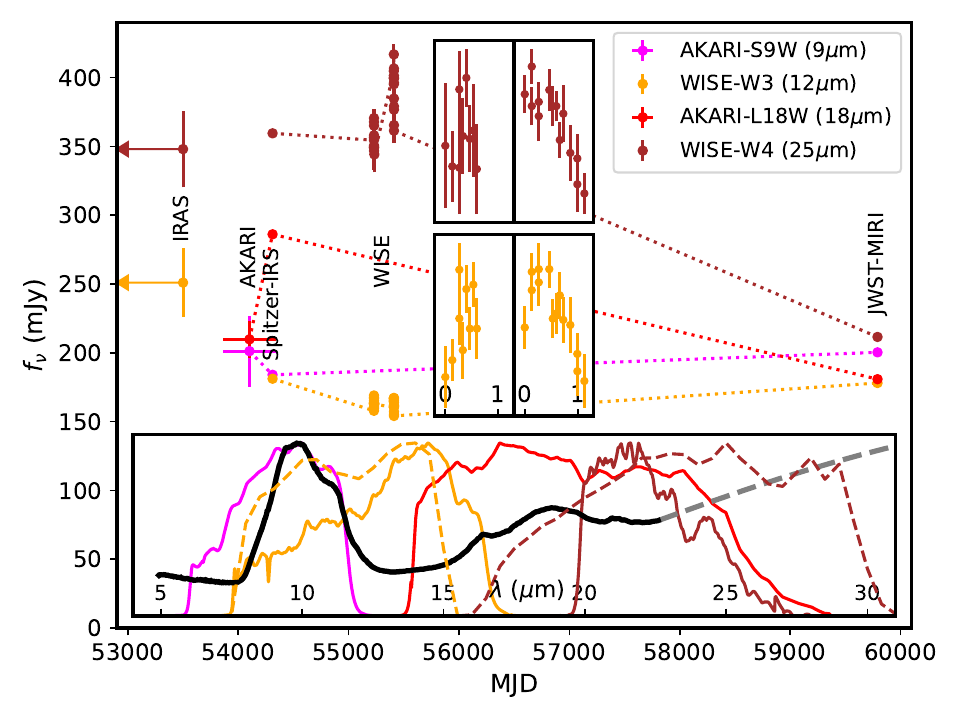}
\caption{Mid-infrared photometry or synthetic photometry (from \spitzer{} IRS and \jwst{} MIRI spectra) of \thestar{} vs time.   \emph{IRAS} observations occurred in 1983 (MJD=45374--45659, off-scale to the left).  The \emph{AKARI} observations occurred during a 477-day all-sky survey represented by the horizontal error bars.  The top and middle insets show details of multi-epoch W3- and W4-band photometry obtained over two $\sim$one-day intervals during the \wise{} cryogenic mission. The \jwst{} spectrum is plotted as the black-line in the bottom inset, with the grey dashed line an extrapolation using the best-fit model in Fig. \ref{fig:sed_fit}.  Colors correspond to \emph{passbands}, not necessarily the telescope, and the response profiles are plotted in the lower inset.  The dashed yellow and brown lines are the \emph{IRAS} 12- and 25-\micron{} profiles.}  
\label{fig:ir}
\end{figure}

We revisited variability in emission at $\lambda \gtrsim 10$ \micron{} with 12- and 25-\micron{} photometry from the \emph{IRAS} Faint Source Catalog based on its all-sky survey \citep{Neugebauer1984,Moshir1990}, 9- and 18-\micron{} photometry from the \emph{AKARI} mission \citep{Ishihara2010}, 12- and 25-\micron{} photometry from the cryogenic \wise{} mission \citep{Wright2010,Cutri2013}, and synthetic photometry in the \emph{AKARI} and \wise{} passbands using the \spitzer-IRS and \jwst-MIRI spectra (Fig. \ref{fig:ir}).  \emph{IRAS} data were obtained 23 years before the next observations, and \emph{AKARI} photometry is based on 3--4 observations over the span of 477 days. \spitzer{} and \jwst{} spectra extend to $\lambda = 22.6 \micron$ and the \wise{} 25-\micron{} channel is incompletely covered, so we extrapolated using a scaled version of the best-fit curve in Fig. \ref{fig:sed_fit}.  (The estimated fluxes are not sensitive to the exact form of the extrapolation.)  

While the flux in the 9-\micron{} \emph{AKARI} passband has been stable since 2006--2007, emission at longer wavelengths show significant variation.  The apparent elevated flux at 12 \micron{} observed by \emph{IRAS} is likely due to a broader bandpass (dashed yellow line in lower inset of Fig. \ref{fig:ir}) that includes the 10 \micron{} silicate emission feature.   The estimated 25-\micron{} flux of \jwst{} is significantly lower than that from \spitzer, echoing the results of \citet{Perotti2023}, and the \spitzer{} emission is consistent with both earlier \emph{IRAS} and later \wise{} photometry.  Moreover this decline is also manifest in estimated 18-\micron{} fluxes, probably because the \emph{ARAKI} bandpass overlaps with the \wise{} 25-\micron{} one (bottom inset of Fig. \ref{fig:ir}).  

In addition, rapid variation ($\sim$0.15 mag over $\sim$1 day) is present in multi-epoch \wise{} W3 and W4 photometry (upper insets of Fig. \ref{fig:ir}).  Emission from these wavelengths is coming from cool ($\lesssim 300$K) dust at distances where the orbital period is $\sim1$~yr and thus the variation is faster than any dynamical timescale.  Shadowing by a variable inner disk wall could explain the short timescale, however, on long timescales, the overall decline at longer wavelengths is not mirrored by a contemporaneous increase in flux at 3.4 or 4.6 \micron{} (as in the case of see-saw variability).  There is an overall increase in emission at these wavelengths over the duration of the \emph{NEOWISE} mission but the most recent released values were obtained only 16 days before the \jwst{} observations, and are similar to those obtained during the \wise{} cryogenic mission, when simultaneously measured 25-\micron{} fluxes were similar to the \spitzer{} values.  One reason for the lack of see-saw variability may be the moderate disk inclination ($\sim$50 deg); inclinations $>70$ may be required for a single fulcrum wavelength to appear in the time-dependent SED \citep{Bryan2019}.     

\subsection{Variability in accretion}
\label{sec:var_accretion}
\begin{figure}
    \centering
    \includegraphics[width=\columnwidth]{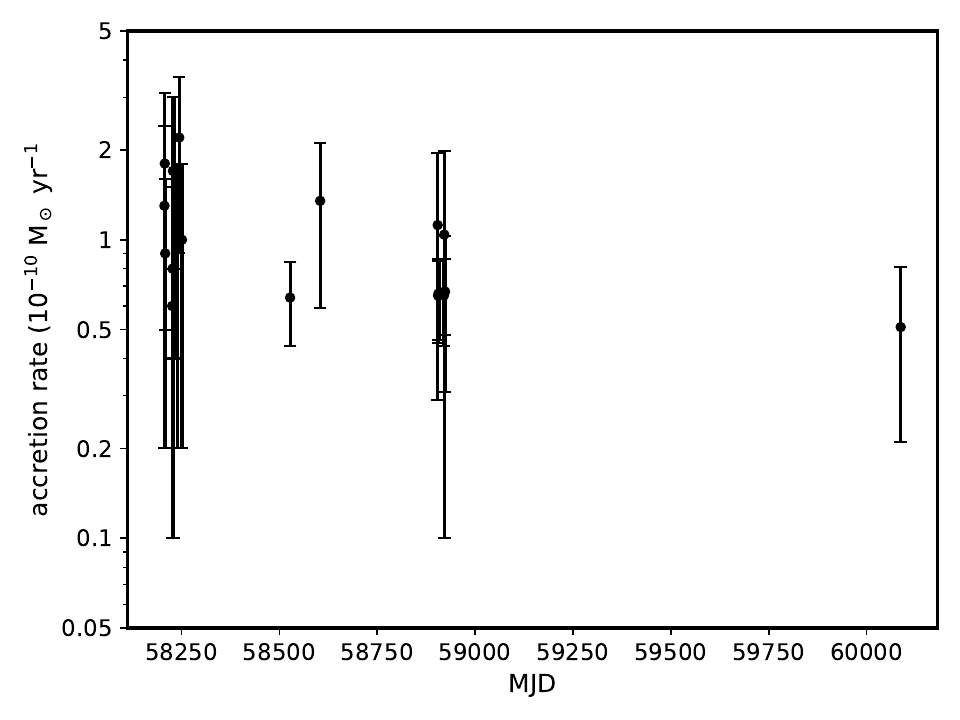}
    \caption{Accretion rate inferred from fitting an magnetospheric accretion model to the H$\alpha$ emission line profile obtained by the HARPS and NRES spectrographs.}
    \label{fig:accretion_rate}
\end{figure}

The accretion rate inferred from model-fitting of the Balmer H$\alpha$ line was consistently below a few times $10^{-10}$ \msun~yr$^{-1}$ over a baseline of nearly three years (Fig. \ref{fig:accretion_rate}), at the lower end of the observed distribution among T Tauri stars \citep{Hartmann1998,Manara2022}.  However, due to the large uncertainties in the values, we are unable to detect significant variation ($\chi^2 = 6.1$ for $\nu = 22$).  The large uncertainties are due to the degeneracy with the gas temperature, which is unconstrained and subject to vary with changes in the magnetospheric accretion radius (see Sec. \ref{sec:synthesis})   

\subsection{Model-based synthesis}
\label{sec:synthesis}

\begin{figure}
\includegraphics[width=\columnwidth]{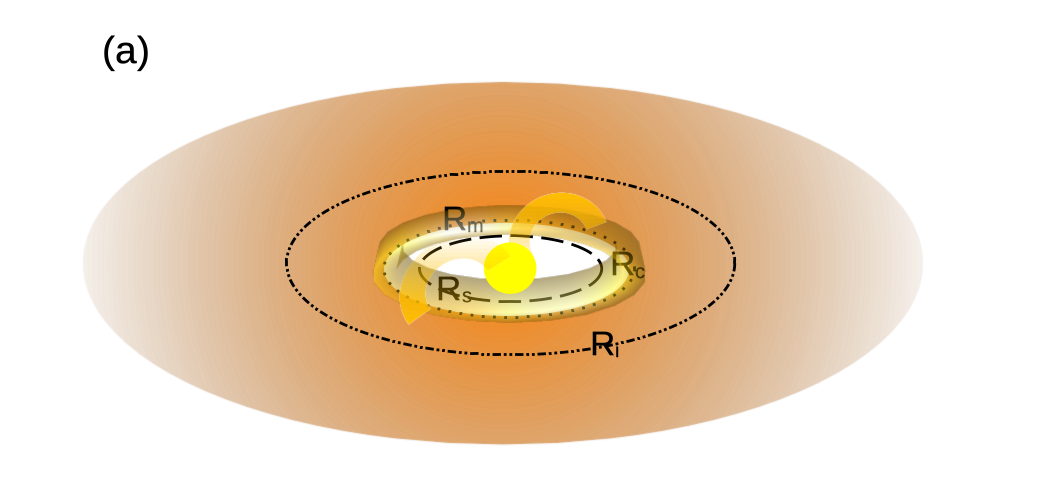}
\includegraphics[width=\columnwidth]{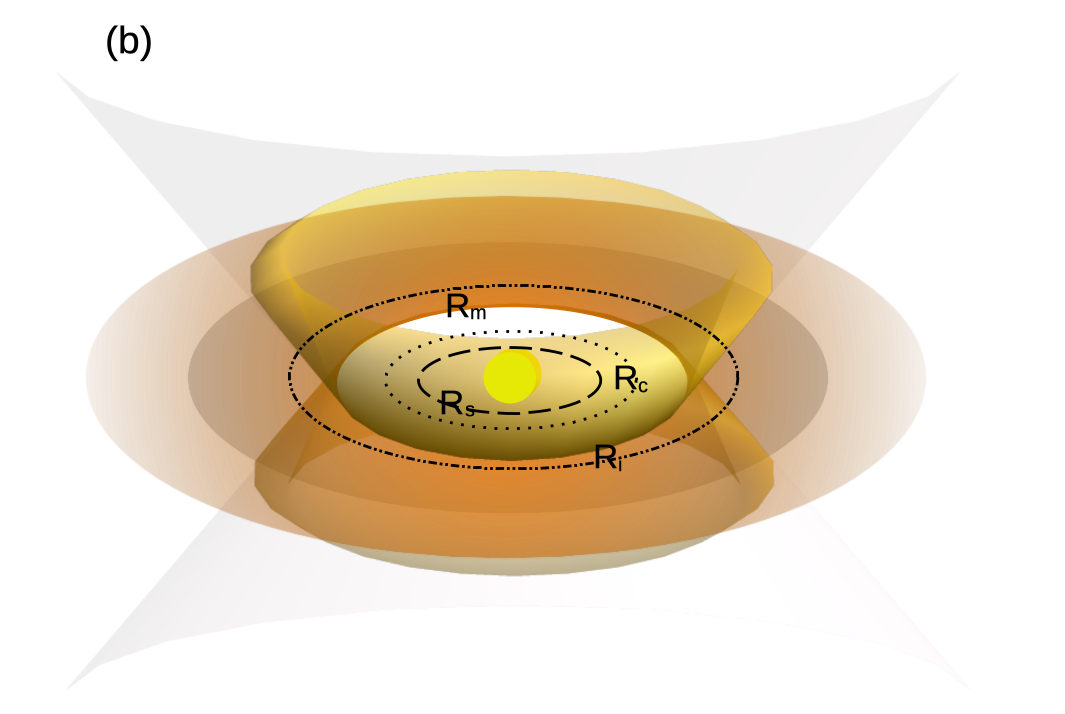}
\includegraphics[width=\columnwidth]{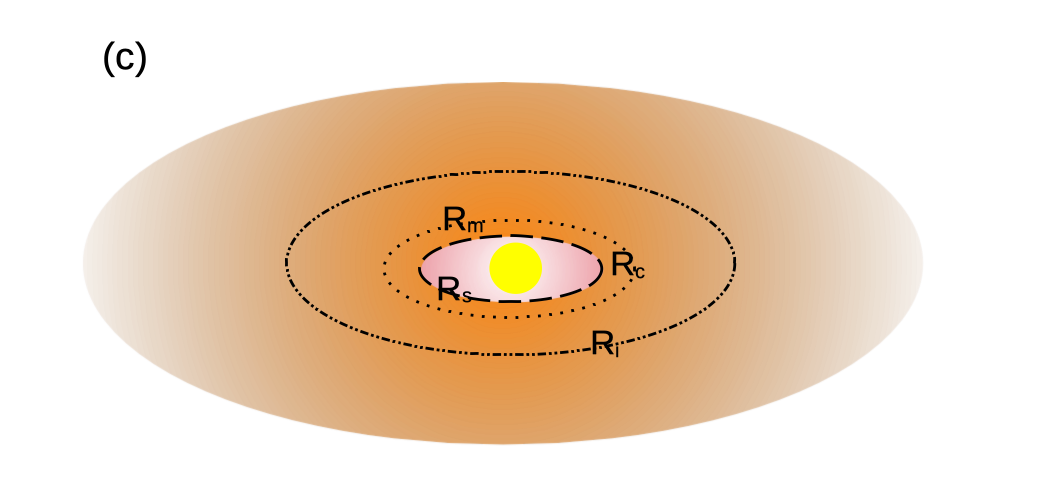}
\caption{Schematic illustrating how variation in the inner disk truncation radius $R_m$, relative to the co-rotation radius $R_c$, dust sublimation radius $R_s$, and gas ionization radius $R_i$ could produce the observed pattern of optical and infrared variability of \thestar{}.  (The viewing geometry is 51 deg inclination and nothing is to scale.) In scenario (a) an elevated disk accretion rate or weak magnetic field means that $R_m$ lies at or interior to $R_c$ and non-axisymmetric magnetospheric accretion occurs via onto the star via streams or ``funnels".  Hot dust in the raised inner edge of the disk is responsible for excess emission at 3.4 and 4.6 \micron{} but the star is only occulted periodically by accretion streams.  In scenario (b) a low accretion rate or stronger magnetic field moves $R_m$ beyond $R_c$ but inside $R_i$, and the disk is diverted into a magnetized dusty wind which produces a large 3.4/4.6 \micron{} excess, partially shadows the outer part of the inner disk, and continuously occults the star.  In scenario (c) an interval of magnetic reversal with a very weak dipole component moves $R_m$ inside $R_s$ so that magnetospheric accretion is dust-free and there is little disk emission at 3.4 and 4.6 \micron.}
    \label{fig:cartoon}
\end{figure}

We describe the dynamic behavior of \thestar{} in terms of the relative location of four locations in the disk: (1) the co-rotation radius $R_c$ inside or outside of which disk loses or gains angular momentum through magnetic coupling to the star; (2) the truncation radius $R_m$ where the disk is vertically diverted by the large-scale stellar magnetic field; (3) the disk ionization radius $R_i$ interior to which the disk is affected by the magnetic field; and (4) the dust sublimation radius $R_s$.  Figure \ref{fig:cartoon} illustrates this behavior (described in more detail below).  \citet{Liffman2020} proposed an analogous scenario to explain the variable infrared emission of the pre-transition disk LRLL~31, i.e., from a dust ``fan" ejected from the inner disk by a jet/wind.  In their model, variation in accretion rate drives variation of the disk truncation radius and the speed of the wind, which in turn controls the height of the dust fan and thus the level of infrared emission. 

If the truncation radius is exterior to the dust sublimation radius, dust that is dynamically bound to the diverted gas could produce both the optical and infrared variability, by scattering along the line of sight, and emission, respectively.    Whether the flow is magnetically ``funneled" onto the star (Fig. \ref{fig:cartoon}a) or diverted into a ``propeller" outflow and wind (Fig. \ref{fig:cartoon}b) then depends on the whether the truncation radius is interior or exterior to the co-rotation radius given by:
\begin{equation}
R_c = 0.02~{\rm au~} R_* M_*^{1/3} P_{\rm rot}^{2/3},
\label{eqn:corotation}
\end{equation}
where $M_*$ and $R_*$ are the the stellar radius and mass in solar units \citep{Ustyogova2006,Bessolaz2008}.  \citet{Takasao2022} carried out three-dimensional magnetohydrodynamic (MHD) simulations of this phenomenon and found that the magnetic truncation radius $R_m$ closely followed the scaling relationship proposed by \citet{Ghosh1979}:
\begin{equation}
\label{eqn:truncation}
    R_m = 0.063~{\rm au~} R_* B_*^{4/7}R_*^{5/7}M_*^{-1/7}\dot{M}_{-10}^{-2/7},
\end{equation}
where $B_*$ is the large scale dipole field evaluated at the stellar surface (in kG), and the accretion rate $\dot{M}_{-10}$ is in units of $10^{-10}$~\mdotyr.   

We adopted a value of 0.44 kG for the large-scale dipole field evaluated at the stellar surface of 22\% of the total field strength as found by \citet{Lavail2019} to be typical (see discussion in \citealp{Gehrig2022}).  The  total field strength was in turn based on a relation with the Rossby number $Ro$, the period rotation \citep[3.03~d,][]{Thanathibodee2020} normalized by the convective turnover time $P_{\rm rot}/\tau_c$, i.e. ``slow" rotator relation in Table 2 of \citet{Reiners2022}.  The convective turnover time $\tau_c$ was estimated as 21.7~d by scaling the solar value with luminosity $L_*^{-1/2}$ \citep{Jeffries2011}, yielding $Ro = 0.16$.  For accretion rate we adopted the range of 0.6--2.2 $\times 10^{-10}$ \mdotyr\ found by \citet{Thanathibodee2020}.    

A third boundary is the radius at which mid-plane temperatures reach $\approx$1000K and alkali metals become partially ionized.  This could also be a transition from region of low turbulent (eddy) viscosity (``dead zone") to a high diffusion region where the magnetorotational instability (MRI) drives turbulence \citep{Desch2015}.  At this point, disk models predict a pressure maximum, and the potential for trapping of solids with sizes such that the Stokes number (ratio of stopping time to orbital time) St will be $\mathcal{O}(0.1)$ \citep{Dzyurkevich2010}.  Depending on grain growth and collisional fragmentation, this could lead to a concentration of dust at the pressure bump and depletion elsewhere in the disk.  The maximum Stokes number (collisional case) is:
\begin{equation}
\label{eqn:stokes_one}
St = \frac{\pi}{\rho_d a}{\Sigma_g}, 
\end{equation}
where $\rho_d$ and $a$ are the internal density and size of the dust, and $\Sigma_g$ is the mass surface density of gas.  Using the relationship for the mass accretion rate in an ``alpha turbulent disk" $\dot{M} = 2\pi \alpha c_s H \Sigma_g$, where $\alpha$ is the turbulent viscosity parameter, $c_s$ the sound speed, and $H$ the vertical scale height of the gas, and the relationship between $H$, temperature $T$, and Keplerian orbital period $P_K$, Eqn. \ref{eqn:stokes_one} can be re-expressed as:
\begin{equation}
\label{eqn:stokes_two}
St \approx \frac{\pi \alpha \rho_d a P_K k_b T}{M\dot{M} m_p \mu}.
\end{equation}
Taking the mid-plane temperature to be 1000K, this results in the scaling:
\begin{equation}
\label{eqn:stokes_three}
St \approx 0.025~\alpha a r^{3/2} \dot{M}_{-10}^{-1},
\end{equation}
where $a$ is in $\micron$ and $r$ is in au.   For $\alpha \sim 10^{-2}$ and $r \sim 0.1$, only cm-sized particles will achieve St $\sim 0.1$ and be trapped.  Thus the dust responsible for the variability and infrared excess is not itself trapped at any dead zone boundary, but it could be produced by fragmentation of trapped particles.  \citet{Ueda2019} explored the conditions for dust trapping, finding a limiting turbulent parameter $\alpha$ in the dead zone that is a function of the dust fragmentation velocity.

We calculated the semi-major axis of dust in radiative equilibrium with the star with temperatures that are below the dust condensation temperature   but above 1000K.  The lower limit is the minimum temperature consistent with the observations (see Fig. \ref{fig:ir}) and is also the temperature at which disk gas starts to ionize (i.e., constituent potassium and sodium) and becomes influenced by the stellar magnetic field.  Condensation temperature varies with disk gas pressure and composition, which also changes depending on what has already condensed out is possibly removed from equilibrium with the gas by gravitational settling, and growth into planetesimals; we adopt values of 1290-1500K depending on the mineral \citep{Lodders2003,Wood2021}. 

Equilibrium temperatures depend on grain size and composition through the absorption (= emission) cross-sections integrated over the stellar SED and 1200K infrared emission (Eqn. \ref{eqn:dust_radius}) using the wavelength-dependent optical coefficients  of \citet{Budaj2015}.  From radiative equilibrium, the dust sublimation radius is
\begin{equation}
\label{eqn:dust_radius}
R_d \approx 0.0047~{\rm au}~R_* \left(\frac{T_*}{T_d}\right)^{2}\left(\frac{\kappa^{\rm abs}_*}{\kappa^{\rm abs}_{IR}}\right)^{1/2},
\end{equation}
where $T_*$ is the stellar effective temperature, and $\kappa^{\rm abs}_*$ is the absorption cross-section averaged over the entire stellar spectrum.  We considered Fe-free forsterite (MgSiO$_2$) and enstatite (MgSiO$_3$) as well as olivine and pyroxenes, magnesium silicates with Fe substituting for 20--50\% of Mg.  The optical absorption cross-sections of the latter are significantly higher due to the presence of Fe \citep{Budaj2015} and their equilibrium temperatures will be higher at the same stellar distance.   

If the accretion rate is low or the magnetic field is strong, then magnetic pressure overcomes disk ram pressure, the truncation radius moves outward to the distance where the disk temperature is 1000K, since below this temperature the disk ceases to be ionized and respond to the stellar magnetic field.  The truncation radius remains at approximately this distance and the flow and resulting IR emission should be relatively stable.   If accretion rates are sufficiently high or the magnetic field is weak, then the magnetic truncation radius can migrate inside the condensation temperature of the dust, and gas diverted above the mid-plane in magnetospheric flow would be dust free (Fig. \ref{fig:cartoon}c).  If magnetically-diverted, dusty accretion flows are responsible for the dipper-like variability of \thestar, then, during this latter, dipping should cease.  Based on the ASAS-SN data (Figs. \ref{fig:all} and \ref{fig:wavelet}) the interval at MJD $\sim$ 57000 would appear to be more quiescent, but the lack of data at earlier times makes this conclusion tentative.

The likelihood that dust will occult the star and the emitting area of its infrared emission, will depend on the vertical distance $z_d$ which dust reaches above the mid-plane.  This is set by the balance of gas drag \citep[i.e., Epstein drag,][]{Li2022} and vertical stellar gravity:
\begin{equation}
\label{eqn:drag}
\rho_g \frac{\pi}{4}a^2 c_s^2 \approx \frac{2 \pi^2 \rho_d a^3 z_d \cos \theta}{3 P_K^2},
\end{equation}
where $\rho_g$ is the local gas density and $\theta$ is the wind half-opening angle \citep[up to 30 deg for a magnetized wind,][]{Blandford1982}.   Assuming axisymmetric flow $\dot{M}$ away from the disk plane from the truncation radius and at the sound speed in a layer that is one gas scale height thick;
\begin{equation}
\label{eqn:gasdens}
\rho_g = \frac{\dot{M}}{2c_s^2 R_m^2 P_K},
\end{equation}
which leads to the relation for the geometric area subtended by the dusty flow:
\begin{multline}
\label{eqn:area}
A = \frac{3}{2\pi} \frac{\dot{M}P_K}{\rho_d a}\\
\approx 0.05 {\rm~au}^2 M_*^{-1/2}\dot{M}_{-10} \left(\frac{R_m}{0.05 {\rm~au}}\right)^{3/2} 
a^{-1} \rho_d^{-1},
\end{multline}
where $\rho_d$ is in g~cm$^{-3}$.  The optical depth of the flow will be 
\begin{equation}
\tau \approx \rho_g H \kappa \epsilon 
\end{equation}
where $\epsilon$ is the dust-to-gas ratio by mass and $\kappa$ is the optical constant averaged over the emission spectrum (e.g. a blackbody at 1200K).  This leads to
\begin{multline}
\tau \approx 0.35 \dot{M}_{-10}\frac{\epsilon}{0.01} \frac{\kappa_*}{10^4 {\rm~cm}^2 {\rm~s}^{-1}} \frac{R_m}{0.05 {\rm~au}} \left(\frac{T}{1000{\rm K}}\right)^{-1/2}\\
\times \bar{\mu}^{1/2},
\end{multline}
where $\bar{\mu}$ is the mean atomic weight of the gas (1.85 for a solar composition).  We calculated the effective emitting area as $A(1-e^{-\tau})$.  

\begin{figure}
\includegraphics[width=\columnwidth]{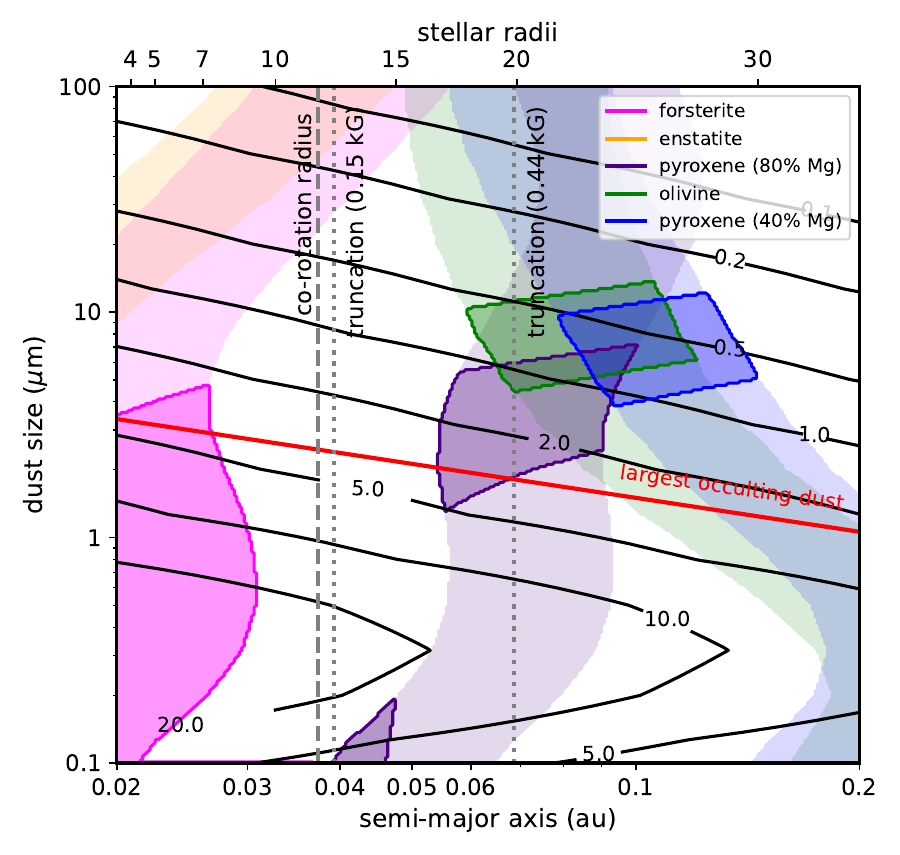}
    \caption{Ranges of dust grain size and semi-major axis that realize equilibrium temperatures between 1000K (the minimum consistent with the observations and the ionization temperature of disk gas) and the $\sim$1300K condensation temperature of the particular mineral.  In the smaller, more heavily shaded regions, dust lofted in magnetically diverted disk gas has an emitting area consistent with values inferred from 3.4 and 4.6 \micron{} photometry (1--4 $\times 10^{-3}$ au$^2$).  Black contours are lines of constant optical depth of the stellar radiation in the Sloan $g$-band due to absorption plus scattering.  The red line is maximum dust size that can reach a height that it occults our line of sight, assuming an inner disk inclination of 52 deg.  The co-rotation radius and magnetic truncation radius of the disk (the latter based on a mean accretion rate of $0.67 \times 10^{-10}$ \msun~yr$^{-1}$ and either a 0.44 kG or 0.15 kG large-scale dipole field) are marked by vertical dashed and dotted lines, respectively.}
    \label{fig:contour}
\end{figure}

Figure \ref{fig:contour} shows regions of semi-major axis $r$ vs. grain size $a$ where the temperature is between 1000--1400K and the effective emitting area is in the range inferred from black-body fitting to the \wise{} 3.4- and 4.6-\micron{} photometry ($1-4 \times 10^{-3}$ au$^{2}$; Fig. \ref{fig:wise});  each region is specific to an assumed dust composition (color-coded).  The positions of the co-rotation radius and the disk truncation radius for two values of the large-scale dipole magnetic field evaluated at the surface are indicated as vertical lines. The total optical depth due to scattering and absorption in the Sloan $g$ band-pass from the same dust structure (integrating over the stellar SED and response function) is shown as the black contours in Fig. \ref{fig:contour}.  Finally, we assumed that the inner disk is co-planar with the outer disk \citep[$51.9 \pm 0.1$ deg;][]{Keppler2019} and with the star's rotation axis \citep[$i = 50\pm8$ deg;][]{Thanathibodee2020} so that occultation along our line of sight does not occur unless the dust reaches a height $H = r \cot{i}$ (red line in Fig. \ref{fig:contour}).  

Large ($>$1 \micron) dust with olivine and pyroxene compositions will have temperatures in the necessary temperature range at the predicted magnetic truncation radius, whereas small grains will remain too cool or will vaporize (Fig. \ref{fig:contour}).  Forsterite and enstatite of any size will be too cool at the relevant distances (Fig. \ref{fig:contour}).  More exotic compositions such as graphite and metallic iron (not shown) can also satisfy the temperature/flux constraints, but these  are expected to readily condense only in a strongly reducing, H-rich disk.  The presence of H$_2$O and CO$_2$ and absence of detectable organic molecules \citep{Perotti2023} indicates oxidizing (low C/O ratio) conditions in the inner disk, more favorable to the condensation of Fe-rich magnesium silicates. 

\section{Discussion and Summary}
\label{sec:discussion}

\subsection{A Disk Wind}
\label{sec:wind}

Optical and infrared spectra of \thestar{} contain evidence for a disk wind, i.e. emission in the forbidden line of \ion{O}{1} at 6300\AA\ \citep{Campbell-White2023} and an inverse P Cygni-like profile with blue-shifted absorption in the metastable triplet of \ion{He}{1} at 1.083 \micron{} \citep{Thanathibodee2019}.  The former is exceptionally high compared to stars with disks with similar accretion luminosity, is highly broadened (FWHM $\approx$90 km~s$^{-1}$) relative to nearly all disks, and slightly blue-shifted (8~km~s$^{-1}$).  This suggests emission from a wind arising 0.1--0.2 au from the star \citep{Campbell-White2023}, a distance that overlaps with our predictions for the disk truncation and dust emission radii (Fig. \ref{fig:contour}).

While red-shifted absorption in either H$\alpha$  or triplet \ion{He}{1} would correspond to gas inside the co-rotation radius accreting onto the star, the combination of emission and blue-shifted absorption and emission in the \ion{He}{1} triplet line is typically produced by a stellar wind \citep{Edwards2006}.  Variable, blue-shifted (-35 to -110 km~s$^{-1}$) and red-shifted ($\sim$140 km~sec$^{-1}$) absorption components are seen in \ion{He}{1} spectra of \thestar{} obtained at three epochs prior to MJD=58900 \citep{Thanathibodee2020,Thanathibodee2022}, indicative of magnetospheric accretion.  After MJD=58900, when the photometric behavior has transitioned from a periodic to aperiodic dipper-like behavior, \ion{He}{1} spectra at two separate epochs at the end of 2020/early 2021 lack any red-shifted absorption and instead have red-shifted \emph{emission} and strong blue-shifted absorption, suggestive of a disk wind \citep{Campbell-White2023}.  This is also coincident with a decrease in the baseline stellar brightness and rise in infrared emission (Fig. \ref{fig:trend}).  A spectrum obtained with the iSHELL infrared spectrograph on IRTF in mid-May 2023 (R. Lee, pers. comm.) shows the return of magnetospheric accretion signatures, contemporaneous with the re-emergence of a weak periodic signal (Figs. \ref{fig:all} and \ref{fig:wavelet}), increase in the baseline optical brightness, and decline in infrared emission (Fig. \ref{fig:trend}).  This is consistent with a dusty wind as the source of both the variable emission and optical emission, although not necessarily the same dust at the same distance along the wind.  We caution that the available spectroscopy has limited rotational phase coverage and non-axisymmetric flow could be also be responsible for the observed changes.  

We propose that the fluorescent UV H$_2$ emission detected by \citet{Skinner2022} also emerges from this wind.  For one thing, a wind would screen the disk itself from the Lyman-$\alpha$ photons that pump such emission, precluding it from being the source.  Such emission is common among classical T Tauri stars, generally arises from the inner ($\lesssim$1 au) disk region \citep{Arulanantham2021,France2023}, and a correlation with the narrow-line component of forbidden atomic emission, i.e. [\ion{O}{1}] \citep{Gangi2023} suggests a connection with winds. Fluorescent H$_2$ emission is rare among weak-lined and low-accreting T Tauri stars; another notable exception is EP Cha \citep[RECX-11,][]{France2012}, which is also a ``dipper" star.

\begin{figure}
    \centering
    \includegraphics[width=\columnwidth]{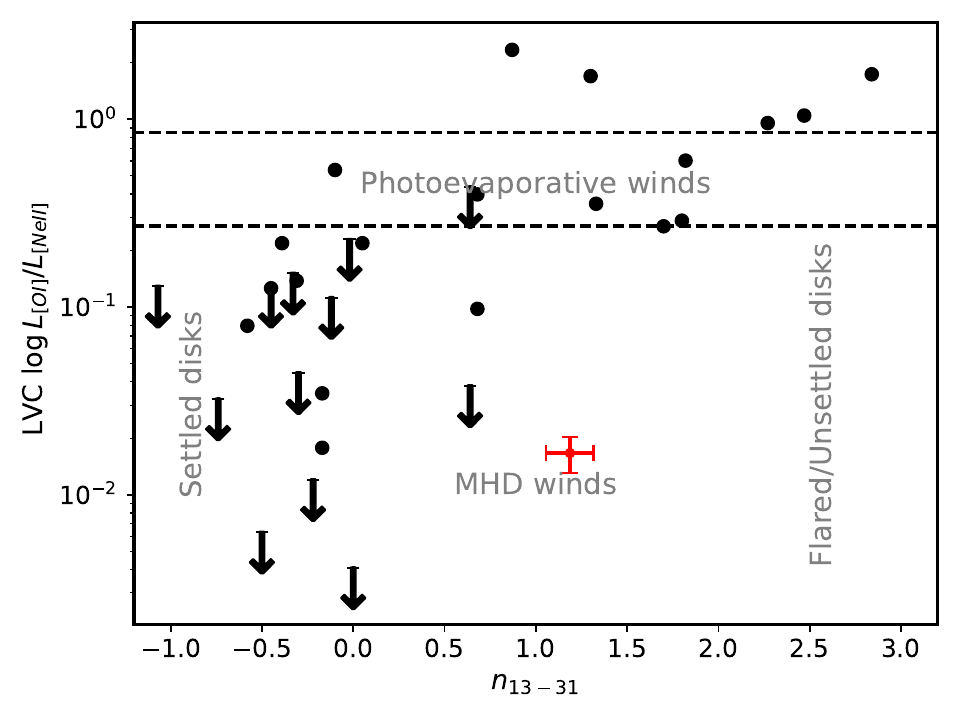}
    \caption{Ratio of the disk-associated, low-velocity component (LVC) of emission in the forbidden lines of \ion{Ne}{2} (12.81 \micron) to that of \ion{O}{1} (6300\AA), which could be an indicator of the presence of a photoevaporative wind vs. an MHD wind, and based on \citet{Pascucci2020}.  Measurements and upper limits from \citet{Pascucci2020} are plotted vs. the infrared spectral index between 13 and 31 $\micron$, which is an indicator of disk evolution, i.e. dust settling and gap clearing \citep{Furlan2009} (based on Fig. 5 in \citet{Pascucci2020}).  The dashed lines bracket the range of ratios expected for photoevaporative winds based on \citet{Ercolano2010}.  \thestar{} is the red point, with the spectral index calculated from the SED in the Fig. \ref{fig:sed_fit}, and the error in $n_{13-31}$ reflecting the difference between the 13 $\micron$ fluxes measured with \spitzer{} IRS and \jwst{} MIRI.  The [\ion{O}{1}] line flux of \thestar{} is from \citet{Campbell-White2023} and the \ion{Ne}{2} line flux is from this work.} 
    \label{fig:oxyneon}
\end{figure}

In addition to the [\ion{O}{1}] 6300\AA{} line, emission in mid-IR forbidden lines of \ion{Ne}{2} (12.81 \micron) and \ion{Ne}{3} (12.55 \micron), whose high ionization potentials mean they can only arise from very hot, escaping gas, are excellent tracers of photoevaporative winds from the disks of T Tauri stars.  \citet{Pascucci2020} has proposed that anemic Ne emission relative to O emission can be explained in terms of an inner MHD wind shielding the disk further out from the hard stellar X-rays that heat the gas (the \ion{O}{1} primarily arising from the MHD wind).   Variability in Ne emission from SZ Cha has been interpreted in the context of a screening wind \citep{Espaillat2023}.  Both Ne lines are detected in MIRI spectrum of \thestar{}, but are weak, and the \ion{Ne}{3} line is blended  with \water{} lines.  The flux in the \ion{Ne}{2} line ($3.5 \times 10^{-16}$ erg s$^{-1}$ cm$^{-2}$, and the red point in Fig. \ref{fig:oxyneon}) is well below that expected for its [\ion{O}{1}] emission based on models of photoevaporative winds \citep[dashed lines in Fig. \ref{fig:oxyneon};][]{Ercolano2010}, presumably due to the presence of the wind at the time of the \jwst{} observation\footnote{Portilla-Revelo et al. (in prep.) find a modestly higher value of $5.7 \times 10^{-16}$ erg s$^{-1}$ cm$^{-2}$ (G. Perotti, pers. comm.).}

\subsection{Timescales and drivers of accretion variability}
\label{sec:timescales}

If the primary source of optical dimming and 3--5 \micron{} emission is dust that is kinematically coupled to magnetized gas in an accretion flow and/or wind from the inner disk, then while variation (i.e., dimming) on the timescale of days can be explained by orbital motion, variation on time scales of months to years, i.e., much longer than the orbital time at the inner disk edge, must be caused by changes in either the geometry of that flow or dust loading of the gas.  In principle, variation in accretion could drive these changes (Eqn. \ref{eqn:truncation}).   We are unable to detect significant variability in accretion rate based on modeling of the H$\alpha$ profile (Fig. \ref{fig:accretion_rate}), but any (detected) variation could also simply reflect changes in the relative amounts of flow through a wind vs. accretion onto the star, rather than the rate through the disk.

At least three timescales are relevant to variable accretion of disk gas \citep{Fischer2022}: the dynamical time scale, which can be as short as $\sim$1 day at the inner disk edge (too short to be relevant), the viscous diffusion time, which is at least $10^4$ times the orbital time (too long), and the thermal time, which is $\sim10^2$ times longer and thus months to years near the inner disk edge (perhaps just right).  In addition, there is a magnetic (Ohmic) diffusion time scale $\tau_m \sim h^2/\eta = \mu_0 \sigma h^2$ \citep[e.g.,][]{Liffman2020} where $\mu_0$ is the permittivity of free space, $\sigma$ the electrical conductivity, and $h$ the relevant scale length, e.g. disk scale height.

In a low-accretion, passively heated disk, disk ionization will depend on irradiation by the star (optical and X-rays) \citep{Jankovic2021}.  
The viscous time scale $t_{\nu} = r^2/\nu$ for a turbulent (e.g., MRI-driven) $\alpha$-disk disk hotter than 1000K is at least a century:
\begin{equation}
t_{\nu} \approx 100 {\rm~yr} \left(\frac{\alpha}{0.01}\right)^{-1}\left(\frac{r}{0.1 {\rm~au}}\right)^{1/2}
\end{equation}
This might create a positive, destabilizing feedback: disk shadowing could lead to suppression of the MRI close to the star, a lower accretion rate, expanded truncation radius, elevated disk rim, and more disk shadowing.  However, again, this cannot explain the year-long variability seen in the optical and infrared.\footnote{This is distinct from any self-shadowing instability of the disk surface \citep{Dullemond2000,Fuksman2022}.}  

Variation in the strength of the large-scale dipole field (i.e., stellar magnetic cycle) could also move the disk truncation radius with respect to the co-rotation radius.  \citet{Armitage1995} showed that magnetic cycles could produce variability in accretion and emission among T Tauri stars, especially in the ultraviolet, and could even be responsible for accretion-driven outbursts \citep{Armitage2016}.  Magnetic cycles are widespread among solar-type stars; little is known about the magnetic behavior of cooler, fully-convective pre-main sequence objects, but \citet{Finociety2023} report variation of a factor of three in the dipole strength of the T Tauri star V1298 Tau over three years.  Full 3-d simulations of fully-convective main sequence M dwarfs (as analogs) suggest that the overall behavior is similar to that of partially-convective stars \citep[e.g.,][]{Kapyla2021,Bice2023b}.   

\citet{Ortiz-Rodriguez2022} found that cycles appear in fully-convective M dwarfs if the magnetic Prandtl number $Pm$ (ratio of turbulent to magnetic diffusivities $\nu_K/\eta$) is $<2$.  The timescale for the reversal is related to that of large-scale diffusion of magnetic fields through the dynamo (convective) region \citep[e.g.,][]{Augustson2015}, and will be of order years, consistent with the few available cases where cycling has been observed for fully convective stars \citep[e.g.,][]{Ibanez-Bustos2019,Ibanez-Bustos2020,Klein2021}.  Generally, more rapidly rotating stars have shorter magnetic cycles \citep{Saar1999,Suarez-Mascareno2016}.  \citet{Bice2023a} found that magnetic cycle time $\tau_{\rm cycle}$ normalized by $P$ in cool dwarf models spanning a large range of parameters scaled as Pm$^{0.35}$Re$^{0.93}$Ro$^{-1.78}$, and is predicted to be $\mathcal{O}(300)$ (days) for Ro $\approx$0.16 (see also \citealt{Strugarek2018}).  Moreover, \citet{LinC2023} find variation in the photometric behavior among 16 T Tauri stars on a timescale of 1.5-4 years.  Thus, cycle times of $\sim$3 yr are plausible, and would be consistent with the time-scale of long-term variation in optical and infrared brightness seen in \thestar{} (Fig. \ref{fig:trend}).  However, if  the time-scale for removal of angular momentum from the local disk is much longer than the magnetic cycling time the disk will only respond to the long-term average dipole field strength.

The disappearance of emission from the disk at 3.4 and 4.6 \micron{} for up to one year at the end of 2014 (around MJD=57000, Fig. \ref{fig:trend}) requires a temporary halt in accretion towards the inner disk edge, a dramatic drop in dust load (e.g., by trapping further out), or a very different geometry of inner disk dust, i.e. one that intercepts much less stellar irradiation close to the star.  The viscous time scale of $\gtrsim10^2$~yr, even at the inner edge, rules out the first scenario. The second is probably precluded by the very small Stokes number of dust and tight coupling of dust to gas in the inner disk (Eqn. \ref{eqn:stokes_three}).   If the disk was truncated (by a stronger magnetic field) at a distance where the temperatures were significantly cooler than the 1000K this would null the 3.4/4.6\micron{} excess, but this would require a significant source of ionizing radiation \citep{Desch2015}.  

Alternatively, dramatic weakening of the dipole field during a particularly prolonged polarity reversal \citep{DeRosa2012} would allow the disk inner edge to migrate interior to the dust sublimation radius.  For example, reduction of the dipole strength by a factor of three (and magnetic energy by a factor of $\sim$10) as seen in the Sun \citep{DeRosa2012}, moves the truncation radius well inside the sublimation radius of Fe-bearing dust and the co-rotation radius of \thestar{} (Fig. \ref{fig:contour}).   Magnetospheric accretion could still occur, but the gas would be dust free and will have low continuum emission.\footnote{Fe-free forsterite dust could survive closer to the star but only in the absence of heating of the gas by dust with other compositions.}  The relatively ``flat" disk geometry exterior to the sublimation radius would intercept and reprocess much less stellar radiation.  It is possible that shorter episodes occur but have been missed due to the 6-month cadence of NEOWISE monitoring . 

\subsection{Composition of the inner disk}
\label{sec:composition}

As opposed to the outer disk, the inner disk cannot be ``primordial":  combining an inner disk gas mass of $7.6 \times 10^{-4}$ \mjup\ \citep{Portilla-Revelo2023} with an accretion rate of $0.6-2.2 \times 10^{-7}$ \mjup~yr$^{-1}$ \citep{Thanathibodee2019} returns a residence time of 3--12 kyr, far shorter than the star's estimated age of $5 \pm 1$ Myr \citep{Keppler2018} \footnote{The PDS 70bc planets, and thus the gap they formed, are likely younger than the star, but it is usually assumed that giant planets have commenced runaway growth and gap opening at a system age of $\sim$1 Myr \citep{Cridland2019}}.  This would be the case even if the gas mass is underestimated due to depletion of CO (which is used to estimate total gas mass) by freeze-out in the outer disk. 

The inner disk must be resupplied across the gap from the outer disk, and/or regeneration of gas/dust from collisions, disruption, or evaporation of planetesimals.  Detection of H (red-shifted H$\alpha$ absorption and H$_2$ emission) and  He (reversed triplet He I line) unambiguously indicate the presence of primordial gas (from the outer disk), possibly via a ``bridge" structure in the gap identified in ALMA imaging of both gas and dust \citep{Keppler2019}.  Partial trapping of condensible solids (including, possibly, CO ice) is expected in the pressure bump at the inner edge of the \emph{outer} disk \citep{Pinilla2017}, and can explain the observed concentration of dust there \citep{Portilla-Revelo2023}.  Thus disk material passing within the gap should be dust-poor, as is observed \citep[dust:gas ratio of $\approx$630,][]{Portilla-Revelo2023}.  

The inner disk itself, however, appears to be gas-poor or dust-rich.  \citet{Portilla-Revelo2023} derived an inner disk gas-to-dust ratio of $\sim$10 (dust-rich relative to ISM) based on relative line ($^{12}$CO) vs. continuum emission at mm wavelengths.  The gas-to-dust ratio along the line of sight to the star (i.e., in the wind) can also be estimated by comparing absorption of X-rays \citep{Wilms2000} to scattering/absorption in visible light by dust.  
\begin{equation}
{\rm gas:dust} \approx \frac{2.5 N_H m_p \bar{\mu} \kappa_m}{\ln 10~\Delta m},
\label{gastodust}
\end{equation}
where $N_H$ is the hydrogen column density inferred from X-ray absorption (assuming a solar-metallicity gas\footnote{X-ray emission from \thestar{} peaks at 1 keV, at which gas and dust make nearly equal contributions to the absorption \citep{Bethell2011}, thus the assumptions of solar metallicity could lead to $N_H$ underestimated by up to a factor of two.}), $m_p$ is the proton mass, $\kappa_m$ is the specific mean opacity of the dust in a optical passband, and $\Delta m$ is the corresponding dimming in that passband.  \citet{Joyce2023} obtained 0.2--12 keV X-ray spectra of PSD 70 with the \emph{XMM-Newton} telescope at 6 epochs spaced a day apart.  They derived a mean $N_H$ of $2.2 \pm 0.2 \times 10^{20}$~cm$^{-2}$ but there is evidence for variability ($\chi^2 = 28.7$ for 5 degrees of freedom).  We set $\Delta g$ = 0.25 mags based on ASAS-SN photometry during the \emph{XMM-Newton} observations, and $\kappa = 2 \times 10^4$ cm$^{2}$~g$^{-1}$ based on Fig. \ref{fig:dustsize}, and derive a gas-to-dust ratio of $\approx$60.  This value applies only along the line of sight to the star and should be considered an upper bound to the disk value because, if the inclination of the inner disk is $\approx$50 deg \citep{Thanathibodee2020}, the line of sight will probe far above the disk mid-plane where depletion of dust due to gravitational settling will be significant.

Freeze-out and trapping of CO by the predicted bump at the inner edge of the outer disk would deplete CO gas sourced to the inner disk, and thus lead to an erroneously low CO-based estimate of gas mass based on a canonical abundance.  However, condensible solids (i.e., dust-forming elements in the inner disk) should be depleted by at least as much, and thus the gas-to-dust ratio should only increase. Moreover, we would expect the growth of planetesimals and planets in the inner disk to sequester solids, increasing the gas-to-dust ratio.  As suggested by \citet{Benisty2021}, evidence for a dust-rich disk thus suggests internal production of dust analogous to that in a debris disk, i.e. by planetesimal collisions, disruption, or evaporation.  

Emission from \thestar{} in infrared \water{} lines falls along an inverse trend with the slope of the continuum emission in the 13-30 \micron{} range from T Tauri stars, explainable by the formation of a gap or central cavity \citep{Perotti2023}.  This emission is dominated by a compact ($\sim 7 \times 10^{-3}$ au$^2$) source at $\sim$600K \citep{Perotti2023}.  The absence of \water{} emission at a temperature corresponding to the $\sim$1200K dust component suggests that \water{} is efficiently sequestered somewhere in the inner disk and does not reach the point where the MHD wind is launched.  The cooler H$_2$O emission could be explained by a wind that shields the disk from the stellar XUV radiation that normally heats the disk atmosphere lying above the mid-IR continuum $\tau = 1$ level \citep{Woitke2018}.\footnote{Although gas-dust thermal disequilibrium is possible, and gas can be cooler than dust in the outer disk \citep{Facchini2017}, models generally predict higher gas temperatures in the inner disk \citep{Kamp2004}.}

There is also \cotwo{} emission is from cool gas.  Photometry by ASAS-SN during \jwst-MIRI observations indicates the system was in the ``dipper" mode (Figs. \ref{fig:all} and \ref{fig:wavelet}) with relatively low emission at 25\micron{} (Fig. \ref{fig:ir}) suggestive of shadowing, which could partially explain the cool emission temperature of CO$_2$.  Importantly, any variability at $\ge 25$ \micron{} (Fig. \ref{fig:ir}) could affect the continuum slope and $n_{13-31}$.  

Although the term ``hybrid disk" seems appropriate for the inner disk of \thestar, this term has already been adopted for disks that are \emph{depleted} in dust relative to the T Tauri phase for a given mass of gas \citep{Pericaud2017,Miley2018}.    Thus we call a dust-rich disk that is comprised of primordial gas that is depleted of dust but \emph{enriched} by secondary debris and gas from planetesimals a ``chimera disk".

\subsection{Future directions} 
\label{sec:future}

Obviously, additional \jwst{} observations of \thestar{} or more advanced analysis of existing observations \citep{Perotti2023} could play an important role in understanding the inner disk of \thestar.  \jwst{} NIRSpec observations of \thestar{} at  0.6-5 \micron{} have already been performed and would provide additional insight into its composition. 

Particularly important in more fully describing the behavior of the inner disk of \thestar{} and testing the scenarios presented here would be long-term, parallel monitoring of indicators of the wind and accretion ([O~I],triplet He~I), dust occultation (optical photometry), and emission in the infrared.  For the latter, while NEOWISE monitoring at 3.4 and 4.6\micron{} is sensitive to changes in the inner disk, its cadence is limited to a day-long visit every six months.  The \emph{AKARI} mission observed the sky at multiple epochs \citep{Tachibana2023} but these data are not yet publicly available.  

\begin{figure}
\includegraphics[width=\columnwidth]{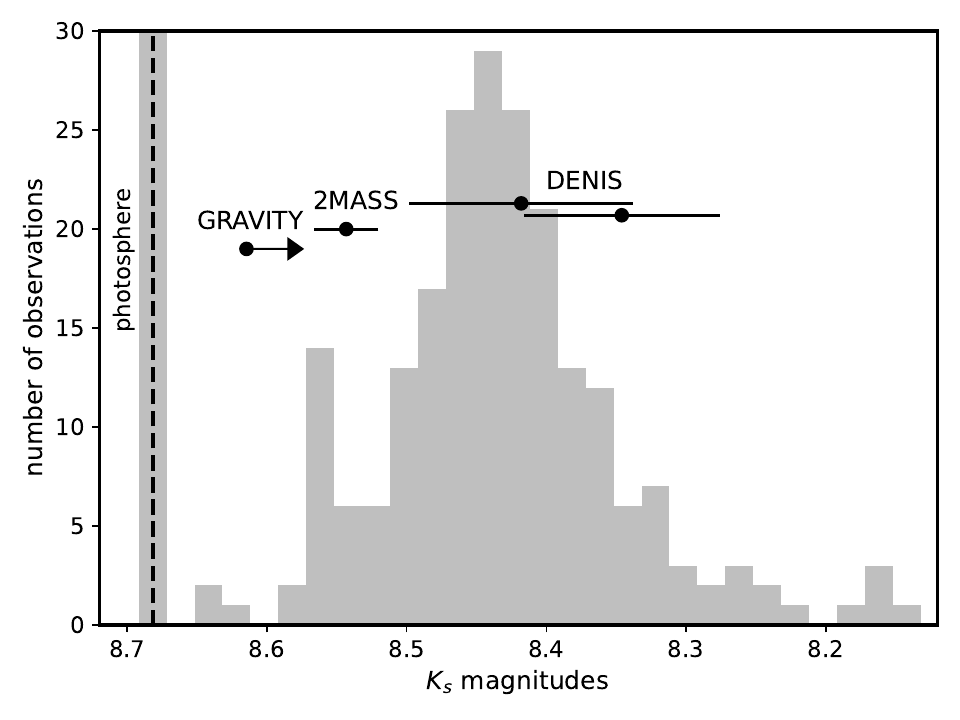}
    \caption{Predicted distribution of the $K$-band brightness of \thestar{} based on black-body fits to the multi-epoch \emph{NEOWISE} 3.4- and 4.6-\micron{} band photometry (Fig. \ref{fig:wise}), binned in 0.02 mag intervals (the approximate limiting precision of ground-based infrared photometry).  Also shown is the 2MASS measurement in 1999 and observations at two epochs from the DENIS survey (3rd release) separated by 100 days in 1998.  The limit from the 2018 VLTI-GRAVITY observations \citep{Wang2021} is the 6\% of the total flux (which must equal or exceed that of the photosphere) that was resolved ($>$0.2~au).  The photosphere value, based on the best-fit of stellar models to data at shorter wavelengths is indicated as the dashed line, and the corresponding bin contains 38 epochs.}
        \label{fig:kmags}
\end{figure}

On the other hand, excess emission from the inner disk is readily detected in the 2.2 \micron{} $K$-band (Fig. \ref{fig:sed_fit}) from the ground.  Fig. \ref{fig:kmags} plots the distribution of $K$-bands magnitudes predicted from the black-body fits performed on the NEOWISE photometry and compares these to 2MASS, two epochs of DENIS photometry, and a lower limit on the value from the resolved component from VLTI-GRAVITY \citep{Wang2021}.  The discrepancy between the 2MASS and DENIS photometry is not due to a difference in the response function since these are very similar.  We propose that the offset of the distribution from the photosphere value inferred from stellar model fitting (vertical dashed line in Fig. \ref{fig:kmags}) is the minor contribution of the disk beyond its inner edge to the $K$-band flux, i.e. a hot dusty upper atmosphere. The distribution shows that time-series monitoring of \thestar{} at 2.2 \micron{} with 2MASS-like precision could successfully track inner disk emission.

The large uncertainty in the accretion rate of \thestar{} inferred from model fits to the H$\alpha$ line profile is due to the degeneracy with gas temperature \citep{Muzerolle2001}. The accretion rate of \thestar{} can be better constrained by modeling the profiles of multiple H~I lines (rather than only the Balmer $\alpha$ line) that form at different temperatures/regions in the accretion flows. The unknown, variable chromospheric emission in H lines also contributes to the uncertainty at the low accretion rates of \thestar{}. High-resolution, simultaneous spectroscopy of the Balmer and Paschen lines are required to better distinguish the chromospheric contribution of each line and break the accretion rate-gas temperature degeneracy. 

High-resolution interferometry in the infrared could resolve the actual inner disk structure of \thestar{} and determine if the emission at shorter wavelengths is coming from a wind or fan.  \citet{Wang2019} resolved half of the photometrically-estimated excess emission from the disk in $K$-band at 0.2-0.5 au scales with the GRAVITY beam combiner on the Very Large Telescope Interferometer (VLTI).  While they attributed the difference to unresolved emission at $\ll$0.2 au, another explanation is variability.  On brighter stars, GRAVITY interferometry can reconstruct simple parameterized images of inner disks, as was done for the case of DoAr 44, which is about one mag brighter than \thestar{} \citep{Bouvier2020}.  GRAVITY+, which will be an upgrade with laser guide-star adaptive optics, could perform similar observations of \thestar{} \citep{Gravity2022}.  Future interferometry in $L$-band \citep[3.5 \micron, e.g.,][]{Laugier2023} would exploit the higher disk emission at that wavelength.       

Measurement of the magnetic field of \thestar, especially its large-scale dipole component, would allow more robust comparison of disk truncation measurements to theory \citep[Eqn. \ref{eqn:truncation},][]{Gregory2016}.  While measurement of Zeeman broadening in unpolarized light provide information on  total field intensity, spectroscopic monitoring of Stokes $V$ (circular) polarization over a full rotation cycle (Zeeman Doppler Imaging or ZDI) can be used to infer the large-scale topology of the field,  direct measurement of the large-scale dipole requires measurements of all four Stokes parameters \citep{Gehrig2022}.  Such studies require high signal-to-noise, are preferably performed in the near-infrared where the Zeeman effect is stronger, and are observationally expensive.  Line broadening by rapid rotation and line veiling also make these observations challenging.  Recent detailed studies, e.g., of V410 Tau \citep{Caroll2012,Yu2019,Finociety2021} and V1298 Tau \citep{Finociety2023}, demonstrate the potential of the suite of high-resolution infrared spectrographs which, when deployed on 8-m or larger telescopes, could perform similar observations of \thestar.  

Alternatively, conditions for trapping of occulting dust near the co-rotation radius, if confirmed, could constrain the strength of the magnetic dipole field, provided the dust particle size and gas density are known \citep[][see also \citealt{Sanderson2023}]{Zhan2019}.  Dust size can be determined by reddening-extinction trends, e.g., Fig. \ref{fig:dustsize} and gas density can be estimated by X-ray observations \citep[e.g.,][]{Joyce2023} or inferred from the accretion rate.

\section{Summary}
\label{sec:summary}

Time-series photometry and spectroscopy reveal the dynamic nature of the inner disk of \thestar, and multi-wavelength observations point to a chimeric composition brought about by the segregation of the inner and outer disk by the system's two giant planets.  We draw the following conclusions or inferences based on these data:  
\begin{itemize}
\item \thestar{} is optically variable on the timescale of day or days due to intervening sub-micron dust located near the inner edge of this disk.
\item The optical lightcurve of \thestar{} evolves between a periodic and approximately symmetric behavior, and stochastic, asymmetric and ``dipper"-like behavior.  This could be due to the changing position of the inner edge of the magnetically truncated relative with respect to the co-rotation radius which alternatively drives magnetospheric accretion vs. ``propellor"-like accretion with a magnetized wind. 
\item The infrared SED of the inner disk of \thestar{} can be modeled as a warm ($T \lesssim 600$K), radially extended disk and a quasi-isothermal component at 1000--1500K, with the latter entirely responsible for emission at 3.4 and 4.6 \micron{}. The temperature range of the hot dust spans the ionization temperature of disk gas and typical dust sublimation temperatures.
\item Variability of \thestar{} at 3.4 and 4.6 \micron{} is produced by changes in the magnetic truncation radius of the disk relative to the co-rotation radius and the location at which disk material is diverted vertically by the field and the extent which it intercepts and reprocesses stellar irradiation.  This variability is pronounced in the case of \thestar{} due to the low disk accretion rate and the proximity of the disk truncation, co-rotation, and dust sublimation radii.
\item At high field strength disk gas is diverted outside the co-rotation radius into a magnetized, non-axisymmetric wind which elevates dust into a ``fan" that partially occults the star along our line of sight and produces episodic dimming as well as excess emission at 3.4 and 4.6 \micron{}.
\item At lower field strength, disk gas is diverted closer to the star, and at/within the co-rotation radius; the occultation ceases and infrared emission is diminished; gas reaching the co-rotation radius is accreted onto the star through funnels that periodically occult the line of sight producing a strongly periodic signal and weaker infrared emission.
\item During magnetic reversals, the weak or absent dipole field means that the disk accretes almost directly onto the star, dust is sublimated before that point, there is little or no occultation or excess emission at 3.4 and 4.6 \micron{}.
\item Variability in disk accretion may be self-exciting as the disk wall shadows the disk further out, cooling it below 1000K and the minimum temperature for ionization and MRI instability.   Accretion could also be governed by dust loading and dynamical interaction with protoplanets in the disk
\item The residence time of primordial, H/He-dominated gas in the disk is $<10^5$~yr, much shorter than the age of \thestar, and thus ongoing accretion in the inner disk is sustained by gas from the outer disk crossing the gap occupied by the two giant planets.
\item Gas crossing the gap from the outer disk is observed to be depleted in solids, including ices, perhaps due to a pressure bump where temperatures are $\lesssim$30K, however we find the inner disk to be enhanced in dust, consistent with sub-mm observations.
\item Gas sourced from the outer disk could be recharged with dust and volatile heavy elements (in the form of \water{} and \cotwo) from a population of planetesimals that are now evaporating, disintegrating, and/or colliding. 
\end{itemize}

Future time-series observations, especially infrared photometry and spectroscopy, promise to more fully elucidate the dynamic behavior of the inner disk of \thestar{}, and the potential relationship with the star's magnetic field.


\section*{Acknowledgments}

EG and AH were supported by NASA Award 80NSSC19K0587 (Astrophysical Data Analysis Program) and NSF Award 2106927 (Astronomy \& Astrophysics Research Program).  JMJO acknowledges support from NASA through the NASA Hubble Fellowship grant HST-HF2-51517.001-A, awarded by STScI. This work made use of observations from the Las Cumbres Observatory global telescope network.  Some of the data presented in this article were obtained from the Mikulski Archive for Space Telescopes (MAST) at the Space Telescope Science Institute. The specific observations analyzed can be accessed via \dataset[DOI]{https://doi.org/	
10.17909/s0zk-rq60}. Funding for the \tess{} mission is provided by the NASA Science Mission Directorate. STScI is operated by the Association of Universities for Research in Astronomy, Inc., under NASA contract NAS 5–26555.  This research has made use of the NASA/IPAC Infrared Science Archive, which is funded by the National Aeronautics and Space Administration and operated by the California Institute of Technology.  This research has made use of the SIMBAD catalog VizieR catalogue access tool operated at the CDS, Strasbourg, France.  This publication makes use of data products from the Near-Earth Object Wide-field Infrared Survey Explorer (NEOWISE), which is a joint project of the Jet Propulsion Laboratory/California Institute of Technology and the University of Arizona. NEOWISE is funded by the National Aeronautics and Space Administration.

%

\vspace{5mm}
\facilities{\tess, \wise, \jwst, ASAS-SN, ATLAS, LCOGT}


\software{{\tt astropy} \citep{astropy2013,astropy2018,astropy2022} 
          }



\appendix

\section{LCOGT relative photometry}
\label{sec:lcogt_data}

In each separate bandpass, time-series photometry of an ensemble of stars was calculated by matrix solution of linear equations relating instrumental magnitudes $m_{ij}^{\rm inst}$ of the $j$th observation of the $i$th star with instrument $k$ (a unique site/enclosure/telescope/camera index), apparent magnitudes $m_i^{\rm app}$, observation zero-points $Z_j$, a second-order atmosphere extinction coefficient $\beta$, and an instrument-specific color-term $\gamma$,
\begin{equation}
\label{eqn:photometry}
    m_{ij}^{\rm inst} = m_i^{\rm app} + Z_j + \beta C_i (X_j - 1) + \gamma_{jk} C_i,  
\end{equation}
where $C_i$ is the stellar color and $X_j$ the airmass of an observation.  We adopted the \gaia\ $B_P-R_P$ color for $C$.   When solving for the coefficients of Eqn. \ref{eqn:photometry}, we only used those sources with peak pixel value $< 2 \times 10^4$ to avoid saturation, and which are associated with \gaia\ stars with fractional variability $\sigma<$0.3\%, where:
\begin{equation}
    \label{eqn:variability}
    \sigma \approx \frac{\sigma_f}{f}\sqrt{\frac{N}{8.86}}
\end{equation}
Here $f$ and $\sigma_f$ are the source brightness and its standard error (\gaia\ {\tt phot\_g\_mean\_flux}) and  ({\tt phot\_g\_mean\_flux\_error}), $N$ is the number of individual photometric observations ({\tt phot\_g\_n\_obs}), and 8.86 \citep{Jordi2010} is the mean number of observations during the brief transit of the star across the detector field of view (during which  the star will not vary significantly).  The variance of each individual star was calculated after solution of Eqn. \ref{eqn:photometry}, and a small (3\%) fraction of the most variable stars was removed from the sample before re-calculating the photometry.  The median error in $Z_j$ was used as a metric to track the quality of the solution.  As more of the most discrepant stars were removed in each iteration, the median error decreased, but as sample size decreased error eventually increased again; we adopted the solution at the minimum median error for calculating light curves for each star.   To achieve successful convergence it was necessary to restrict reference stars to those with similar \gaia\ $G$-magnitude and $B_P-R_P$ colors.  



\clearpage
\newpage



\end{document}

%% file: newcommands.tex
\newcommand{\halpha}{H$\alpha$}

\newcommand{\water}{H$_2$O}

\newcommand{\cotwo}{CO$_2$}

\newcommand{\tess}{\emph{TESS}}

\newcommand{\gaia}{\emph{Gaia}}

\newcommand{\jwst}{\emph{JWST}}

\newcommand{\spitzer}{\emph{Spizter}}

\newcommand{\wise}{\emph{WISE}}

\newcommand{\msunyr}{\rm{M_{\sun} \, yr^{-1}}}

\newcommand{\msun}{M$_{\odot}$}

\newcommand{\mjup}{M$_{\rm JUP}~$}

\newcommand{\mdotyr}{$M_{\odot}$~yr$^{-1}$}

\newcommand{\teff}{\ensuremath{T_{\rm eff}}}